\documentclass[fleqn,usenatbib]{mnras}

\usepackage{newtxtext,newtxmath,orcidlink}
\usepackage{natbib}

\usepackage[T1]{fontenc}
\usepackage{mathtools}
\usepackage[thinc]{esdiff}

\DeclareRobustCommand{\VAN}[3]{#2}
\let\VANthebibliography\thebibliography
\def\thebibliography{\DeclareRobustCommand{\VAN}[3]{##3}\VANthebibliography}

\usepackage{graphicx}	
\usepackage{amsmath}	
\usepackage{subcaption}
\usepackage{booktabs}
\usepackage{xcolor}

\newcommand\rj{{\,{\rm R}_{\rm J}}}
\newcommand\mj{{\,{\rm M}_{\rm J}}}
\newcommand\mearth{{\,{\rm M}_{\oplus}}}

\def\msun{{\rm M_{\odot}}}

\title[FFPs ejected from self-gravitating discs]{Disc fragmentation. I. Ejection of Jupiter-mass Free Floating Planets from growing binary systems}

\author[Aleksandra \'{C}alovi\'{c}]{Aleksandra \'{C}alovi\'{c}$^{1}$\thanks{Contact e-mail: \href{mailto:ac874@le.ac.uk}{ac874@le.ac.uk}}, Sergei Nayakshin$^{1 \orcidlink{0000-0002-6166-2206}}$, Sarah Casewell$^{1}$, Núria Miret-Roig$^{2, 3}$
\\
$^{1}$School of Physics and Astronomy, University of
  Leicester, Leicester LE1 7RH, UK\\
$^{2}$Dep.~de Física Quàntica i Astrofísica (FQA), Univ.~de Barcelona (UB), Martí i Franquès, 1, 08028 Barc., Spain\\
$^{3}$Institut de Ciències del Cosmos (ICCUB), Univ.~de Barcelona (UB), Martí i Franquès, 1, 08028 Barcelona, Spain}

\date{Accepted XXX. Received YYY; in original form ZZZ}

\pubyear{\the\year{}}

\begin{document}
\label{firstpage}
\pagerange{\pageref{firstpage}--\pageref{lastpage}}
\maketitle

\begin{abstract}
Over the past 25 years, observations have uncovered a large population of free-floating planets (FFPs), whose origins remain debated. Massive FFPs (several Jupiter masses or more) may form via gravitational collapse of molecular clouds, similar to stars. Lower-mass FFPs likely originate in planetary systems and are later ejected through dynamical interactions. We show that disc fragmentation in very young stellar binaries may be an abundant source of  Jupiter-like FFPs (JFFPs), with masses above $0.3\,\mj$. In our model, disc fragmentation at tens to 100 au from the primary star produces gas giants, while fragmentation further out forms a more massive object  that will eventually evolve into the secondary star. We present 3D simulations of massive, self-gravitating discs with embedded Jupiter-mass planets and a secondary seed (5–50 $\mj$). Chaotic migration leads to frequent planet–secondary interactions, imparting velocity kicks via gravitational slingshot that usually  end in planet ejection. The ejection fraction increases steeply with  the secondary-to-primary mass ratio, $q_{\rm s}$, reaching $\sim 2/3$ for $q_{\rm s} \gtrsim 0.05$. Compared to Core Accretion JFFPs, disc fragmentation JFFPs: (i) form earlier, and may be more abundant in young clusters; (ii) are ejected at much lower velocities; (iii) may retain compact circum-planetary discs. To reproduce the observed abundance of JFFPs, disc fragmentation planets must be in the post-collapse configuration. They must also either form more frequently in binary systems than around single stars, or be frequently disrupted in inner disc regions, as previously suggested in certain versions of the theory.

\end{abstract}

\begin{keywords}
planets and satellites: formation -- planets and satellites: gaseous planets -- binaries: general
\end{keywords}



\section{Introduction}\label{sec:introduction}

Free-Floating Planets (FFPs) are planetary mass ($M_{\rm p} \leq 13 \mj$) objects that are not bound to any star \citep[e.g.,][]{Yee-25-FFPs-MFunction}. Two main techniques are used to detect these planets: gravitational microlensing \citep[e.g.,][]{Mao_Paczynski_91,Gould_Loeb_92,SuzukiEtal16,Mroz-23-Microlensing-planets-review} and infrared spectroscopic observations of  nearby star forming regions and associations \citep[e.g.,][]{Zapatero_00_JFFP_discovery,Lucas_00_FFP_discovery,Lucas2001-FFP-Orion,Luhman-2004-BD-Chamaeleon,Luhman-16-NGC1333}. The former technique can probe planet masses as low as $M_{\rm p} \sim 0.3\mearth$ \citep{Mroz_20_FFP_Earth_mass,sumi2023}, but requires statistical analysis to constrain planet mass \citep{Jung_24_muFFP_planet}. Infrared observations are suitable for detecting much more luminous, and hence more massive, objects only. Such observations yield planet spectra and luminosity, if the distance to the planet is known. Theoretical models \citep[e.g.,][]{BurrowsEtal97,BaraffeEtal03} can then constrain planet mass but require knowledge of the age of the planet.

As with star-bound planets \citep[e.g.,][]{WF14}, FFPs can be loosely divided on two main varieties: low mass objects, here defined as planets with masses around and below that of Neptune, probably dominated by solid cores; and gas giant planets, which we define as planets with masses in the range from $0.3\mj$ (Saturn mass) to $13 \mj$. We refer to these FFPs as Jupiter-like FFPs (JFFPs) below. \cite{Mroz-23-Microlensing-planets-review} note that three independent long-running microlensing surveys \citep{Gould-22-microlensingFFPs,sumi2023,Mroz-23-Microlensing-planets-review} find similar abundance and mass function of microlensing FFPs ($\mu$FFPs hereafter). In terms of the overall number of objects, $\mu$FFPs are dominated by the lowest masses these surveys are sensitive to, with the mass function roughly following $dN/dM_{\rm p}\propto M_{\rm p}^{-2}$. Although the exact origin of these mainly small mass objects remains debated, it is likely that they formed in stellar binaries\footnote{While single star planetary systems with several gas giant planets can be very effective in ejecting planets \citep{weidenschilling1996,Rasio_96_FFPs,juric2008,Chatterjee_08_PP_scattering}, there are simply too few bound gas giants in observed stellar systems \citep{Fulton_21_giants_occurence} for this channel to work in practice \citep{veras2012,Coleman-24-FFPs}.} and were then ejected by close scatterings with one of the stars \citep[e.g.,][]{Holman_Wiegert_99,PierensNelson08,Coleman-23-planet-in-CB-systems,Coleman-24-FFP-simulations}. 

At Jovian mass and above, the $\mu$FFP population joins the sub-stellar mass end of the IMF of stars \citep[e.g.,][]{Chabrier_05_IMF,Yee-25-FFPs-MFunction}. The exact nature of this transition is yet unclear. The shape of the IMF below $\sim 0.3\msun$ remains hotly debated, with some authors \citep[e.g.,][]{Kirkpatrick_19_BD_IMF,Chabrier_23_IMF} finding approximately log-normal IMF  proposed by \cite{Chabrier_05_IMF}. Others find differing IMF shapes at low masses in nearby star forming regions \citep[see references in][]{Palau_24_BD_formation_review}, implying that the shape of the sub-stellar IMF may be sensitive to environment \citep{Scholz_12_IMF_NGC1333,Kirkpatrick_24_IMF}. For example, \cite{miret-roig2022,Miret-Roig2023-FFP-Review} find a clear excess of massive FFPs in Ophiucus and Upper Sco regions, equating to around 0.04 JFFPs per field star (note: they are only sensitive to JFFP masses $> 4 \mj$). \cite{Yee-25-FFPs-MFunction} reviewed the literature on the mass function in the planetary and sub-stellar regime, finding that, within the current observational uncertainties, there are $\sim 0.1$ to $\sim 0.3$  JFFP in the mass range $0.3\mj$ to 13 $\mj$ per field star.

From the theoretical point of view, the formation channel(s) of sub-stellar mass objects is unclear. The biggest uncertainty is whether the objects form via a star-like gravitational collapse of a molecular cloud, or via a planet-like channel in a protoplanetary disc with ensuing ejection from the system \citep[for a recent review of the topic, see][]{Palau_24_BD_formation_review}. Further, within the star-like formation pathway, two main scenarios can be identified: (i) JFFPs and BDs are much less massive than stars because the clouds from which they form have lower masses than clouds from which stars form \citep[due to  strong MHD driven turbulence, e.g.,][]{Padoan_02_BD_turb_fragm,Hennebelle_08_turb_ISM_theory,Haugbolle_18_turb_star_formation}; (ii) alternatively, the initial cloud mass is in the stellar regime, but the objects were unable to accrete gas from the cloud for very long \citep{Bate_12_cluster_RHD} because of extreme dynamical effects, e.g., ejection from high density structures and discs \citep[e.g.,][]{Reipurt_Clarke_01_BDs,Bate_02_BDs,Stamatellos_07_BDs,Reipurth_15_BD_binaries}, or strong irradiation from nearby OB stars \citep[e.g.,][]{WZ_04_BD_formation}. It is further possible that in realistic star-cluster formation more than one of these scenarios operate and even re-inforce one another \citep[e.g.,][]{Bate-19-Star-Cluster-formation}.

In this paper, we focus on the scenario of forming JFFPs in massive fragmenting self-gravitating protoplanetary discs via ejection by the seed of a growing binary star\footnote{We will present simulations of less massive FFP formation in a follow-up paper, Nayakshin et al 2025}. In the picture we envisage, the formation of both gas giant planets and the secondary star occurs via gravitational instability of the protoplanetary disc \citep[GI hereafter;][]{Kuiper51b,Boss98,Boley09,cha2011,ForganRice13b,HelledEtal13a,KratterL16}. We expect  planet-mass objects to generally form earlier and at closer separations compared to that of the secondary star (cf. \S \ref{sec:general_setup}). Similar to planets growing in binary systems in the framework of the Core Accretion scenario \citep{PierensNelson08,Coleman-23-planet-in-CB-systems,Coleman-24-FFP-simulations,Coleman-24-FFPs}, the secondary star may provide very strong kicks to the planets via close interactions, ejecting them out \citep{Holman_Wiegert_99}. Our goal here is to use 3D numerical simulations of massive self-gravitating discs with embedded planets and a seed secondary to constrain the efficiency of this process, and to seek observables that can distinguish the two competing (CA and GI) planet-like FPP formation scenarios.

The paper is structured as follows: in \S \ref{sec:methods} we introduce our  methodology, starting with the physics setup of the model (\S \ref{sec:general_setup}), the numerical methods (\S \ref{sec:numerical_approach}), and initial conditions of our simulations (\S \ref{sec:initial}), where we also identify our ejection criteria. In \S \ref{sec:results} we present single embedded object experiments (\S \ref{sec:single_planet}) to exemplify their disc migration behavior. We then focus on a few selected planet-secondary and planet-planet interaction examples (\S \ref{sec:examples_of_interactions}), before moving onto the population of the ejected and surviving planets as a whole in \S \ref{sec:ejections} and \S \ref{sec:survivors}, respectively. In \S \ref{sec:analytic} we present an order-of-magnitude analytical explanation of our numerical results, consider the properties of planets that are likely to survive the ejection process, and also estimate the size of circum-planetary discs that they could carry with them.
A broad discussion of our results, including their observational implications and comparison to  previous work, is given in \S \ref{sec:discussion}. We conclude in  \S \ref{sec:conclusions}.

\section{Numerical Methodology}\label{sec:methods}
\subsection{Giant planet and binary formation by disc fragmentation}\label{sec:general_setup}

Disc fragmentation due to GI is understood to occur at wide separations (tens of au or more), since the disc needs to be both massive, and cool rapidly \citep[e.g.,][]{gammie2001,Rafikov05,Clarke09}. However, the details of the fragmentation process and the eventual fate of the fragments remain poorly understood \citep[e.g.,][]{HelledEtal13a,KratterL16}. For example, a number of authors suggest that the initial mass of fragments, $M_{\rm init}$, must be at least $\sim 5\mj$ or more \citep[e.g.,][]{SW09b,KratterEtal10,ForganEtal15-GI-fragments,Xu_25_RHD_disc_fragmentation}. Others find lower fragment masses \citep[e.g.,][]{BoleyEtal10,meru2015}. Simulations that include magneto-hydrodynamics \citep[MHD; ][]{Deng20-MHD-GI,Deng_21_GI_Neptunes,Kubli-GI-MHD-23} yield $M_{\rm init}$ as low as Neptune mass ($\sim 0.05 \mj$). 

In any event, since the disc aspect ratio, $H/R$, increases with radius, $M_{\rm init}$ increases with $R$ \citep[e.g., Fig. 3 in][]{KratterL16}. It is hence likely that GI will form lower mass fragments at tens to $\sim 100$~au, whereas the more massive seed of a binary star would form at wider separations, $R \gtrsim 100$~au. Furthermore, fragments are expected to grow rapidly by gas accretion {\em if} gas can cool rapidly \citep[e.g.,][]{AyliffeBate09b,AyliffeBate12,nayakshin2017}, which again favors large radii as the preferred location for secondary star formation. 

In this paper, based on previous work summarized in \cite{Nayakshin_Review}, we propose the following scheme for coeval formation of JFFPs and binaries. In the Discussion section we shall consider whether this, and other variants of the disc fragmentation theory, could explain JFFPs observations.

\begin{enumerate}
    \item Simulations of protostar's growth from pre-stellar core collapse  \citep{VB05,VB06,VB10,machida2010,MachidaEtal11,VB15} show that protoplanetary disc radius increases with time because the specific angular momentum of material falling later is larger. It is hence likely that disc fragmentation first occurs at smaller radii, hatching planetary mass objects.
    \item Later, when the disc size and mass increase sufficiently, a more massive fragment capable of growing by gas accretion into a massive BDs or a low mass star appears at larger radii.
\end{enumerate}

Numerical simulations show that fragments with low initial masses, $M\lesssim$ a few $\mj$, tend to migrate inward and do not grow in mass significantly, whereas more massive objects accrete gas more rapidly than they migrate, opening a deep gap in the disc and ``running away" in mass to $M \gtrsim 0.1\msun$ \citep{Nayakshin17a}. Our goal below is to detail dynamical evolution of a system that went through both steps introduced above.

\subsection{Numerical approach}\label{sec:numerical_approach}

First-principle hydrodynamical simulations of fragmenting protoplanetary discs \citep[e.g.,][]{forgan2011,ZhuEtal12a,mercer2020} are unfortunately too model dependent on microphysics and computationally expensive to run for relevant time scales (up to $10^5$ years). Here we  follow a pragmatic approach, sidestepping the most uncertain parts of the problem -- the disc fragmentation phase and the transition from relatively low density pre-collapse gas fragments to higher density post-collapse objects \citep[e.g.,][]{Bodenheimer74,HelledEtal08}. As motivated in \S \ref{sec:general_setup}, in the beginning of our simulations, we envisage a number of gas giant planets to be embedded in a massive gaseous disc at tens of au, along with a more massive object at a wider separation. We recognize this is a significant assumption as it assumes planet formation via GI is efficient; however, for now we adopt it as an initial condition and plan to incorporate the formation process more thoroughly in future works.

To simplify the problem, we assume that there is no further disc fragmentation, and that object growth via gas accretion is ineffective. Our simulations focus on object-disc and object-object interactions in the environment of massive post-fragmentation disc. We assume that these interactions are dominated by gravity, and we use a simplified prescription to model disc radiative cooling. The simulations are conducted using 3D smoothed-particle hydrodynamics (SPH) code \textsc{Phantom} \citep{price2018phantom}.


In the the widely used `$\beta$-cooling' model for self-gravitating discs \citep[e.g.,][]{gammie2001, rice2005}, the specific internal energy of an SPH particle, $u$, evolves in time according to
\begin{equation}
    \diff{u}{t} = - \frac{u}{t_{\rm cool}}\;,
    \label{beta-cooling0}
\end{equation}
where $t_{\rm cool}$ is given by
\begin{equation}
    t_{\rm cool} = \frac{\beta}{\Omega}\;, 
    \label{beta0}
\end{equation}
and $\Omega = (G M_*/R^3)^{1/2}$ is the Keplerian angular frequency around the star of mass $M_*$.
The simplicity of the cooling model is advantageous for the analysis of the results, and led to important conclusions about the nature of self-regulated discs and the statistics of perturbations \citep[e.g.,][]{gammie2001,CossinsEtal09,Paardekooper12-Stochastic-Fragm}. Following \cite{nayakshin2017} we use a modified version of eq. \ref{beta0} for disc cooling time:
\begin{equation}
    t_{\rm cool} = \frac{\beta}{\Omega} \Bigg{(} 1 + \frac{\rho}{\rho_{\rm crit}} \Bigg{)} 
    \label{tcool_sn}
\end{equation}
where $\rho$ is the gas density and $\rho_{\rm crit}$ = $10^{-11}$ g/cm$^3$. This modified implementation prevents unphysically fast contraction of optically thick fragments. At large $\beta$ that we employ here, isolated discs are unlikely to fragment. However, massive planets embedded in such discs could potentially promote fragmentation \citep{meru2015}. As is well known \citep[e.g.,][]{Nayakshin10a}, masses and densities of protoplanets formed by GI fragmentation are comparable to masses of first cores in star formation \citep{Larson69}; first cores form at densities exceeding $\sim 10^{-13}$~g cm$^{-3}$ \citep{Larson69}, and their cooling slows down as they become optically thick \citep[][]{masuga200}. The density $\rho_{\rm crit}$ that we choose in eq. \ref{tcool_sn} is almost never reached in our discs, since it exceeds the gas tidal density over much of the disc. We impose a minimum temperature in the disc, set to 10~K.


We use the sink particle prescription \citep[e.g.,][]{Bate95} to account for gas accretion, removing SPH particles that get too close to the planets. We set sink accretion radii to sufficiently small values to minimize gas accretion onto sinks yet to prevent simulations from stalling due to accumulation of gas on very close orbits in circum-planetary disc (which we do not resolve here) around the sinks. We experimented with a variety of values for the cooling parameter $\beta$, from 4 to 30, and determined that values of $\beta$ = 20 -- 30 are best suited for our goals, keeping the discs in a self-gravitating yet not fragmenting state \citep[e.g.,][]{gammie2001,rice2005}. The use of the simplified $\beta$-cooling prescription does not affect generality of our results. In Nayakshin et al 2025 (paper II), we use a self-consistent temperature and density dependent cooling scheme \citep{Marzari_12_circumbinary_discs,ZhuEtal12a} in 2D simulations, and find very similar results.




\subsection{Initial conditions}\label{sec:initial}
\subsubsection{Gas disc}

A spurious disc fragmentation can ensue if initial conditions are not chosen carefully \citep[e.g.,][]{Paardekooper12-Stochastic-Fragm}. The standard way of avoiding this in simulations is to construct an initial disc profile with properties somewhat close to the disc fragmentation boundary and then allow it to relax towards the self-regulated gravito-turbulent state \citep[e.g.,][]{gammie2001,baruteau2011,MalikEtal15}. To do so, we set up an initial disc with no planets, assuming a central star mass of 0.5 $M_\odot$ with a surrounding gas disc comprised of $n_{\rm p}$ = $10^6$ SPH particles, with an inner and outer disc radius of $R_{\rm in}$ = 5 au and $R_{\rm out}$ = 100 au, respectively. The initial surface density profile of the disc is given by
\begin{equation}
    \Sigma(R) \sim \Big{(} \frac{R_{\rm ref}}{R} \Big{)} \exp \Big{[} {-\Big{(} \frac{R}{R{\rm c}} \Big{)}} \Big{]} \Big{(} 1 - \sqrt{\frac{R_{\rm in}}{R}} \Big{)},
    \label{eq:sigma_init}
\end{equation}
where $R_{\rm ref}$ is a reference radius, set as 10 au, $R_{\rm c} =140$ au is the characteristic radius of the exponential taper. The temperature and $(H/R)$ profiles are given by, respectively,
\begin{equation}
    T(R) = 128 \text{ K } \Big{(} \frac{R}{R_{\rm ref}} \Big{)}^{-0.5}\;,
    \label{eq:temp}
\end{equation}

\begin{equation}
    \Big( \frac{H}{R} \Big) =  0.1\; \Big( \frac{R}{R_{\rm ref}} \Big)^{0.25}\;.
    \label{eq:h_r}
\end{equation}

The normalisation in eq. \ref{eq:sigma_init} is set by choosing a minimum Toomre $Q$ parameter \citep{toomre1964}, $Q_0$, over the initial disc. $Q$ is defined as
\begin{equation}
    Q = \frac{c_{\rm s} \Omega}{\pi G \Sigma}
    \label{toomre_q}
\end{equation}
where $c_{\rm s}$ is the sound speed, $\Omega = \sqrt{G M_{*}/R^3}$ the Keplerian angular frequency and $\Sigma$ is the surface density of the disc. For our fiducial runs, $Q_0=$ 1.2, with an initial total disc mass is $M_{\rm D}$ = 0.16 $M_\odot$. 

The simulation is then run for around 6,000 years to allow the initial  small scale SPH particle density inhomogeneities to decay away. At the end of the relaxation stage, disc parameters reach a quasi-steady state. The planets are then added to the disc, and the simulation is restarted and run for an additional 95,000 years to study evolution of the system.



\subsubsection{Adding planets into the disc}\label{sec:add_planets}

At the beginning of the simulations, we simultaneously embed 9 objects on initially circular orbits in the disc, of which 8 are planets with masses 1 -- 3 $\mj$, and one is a secondary object with a mass ranging from zero (no secondary), to masses of $M_2 = $ 5, 8, 10, 15, 25 and 50 $\mj$; the initial configuration of objects and their masses is given in Table \ref{tab:inital}. We assume zero orbital inclination of the planets in the disc; more complicated situations will be presented in future work. This large number of planets added to the simulation domain is not unreasonable based on the physics of fragmenting discs (see below) and is additionally motivated by economy of numerical resources (our typical simulation requires 3 weeks of physical time for a parallel run on 28 CPUs). We find that planet-planet interactions are a minor perturbation to the statistical outcomes of the simulations in terms of planet ejections, our primary interest in this paper. The planets are usually (but not always, see \S \ref{sec:examples_of_interactions} below) in a test particle regime where they do not interact with one another strongly. Having more planets in the disc thus effectively increases the number of numerical experiments we perform at nearly the same computational expense as a one planet plus one massive object simulation. In Appendix \S \ref{sec:Appendix_planet_number} we show that our main results are not affected by choosing 8 planets per disc. Fundamentally, this is because planet-disc and planet-secondary interactions are so much stronger than planet-planet interactions: in our simulations, the disc and the secondary are $\sim 2$ orders of magnitude more massive than a planet.

The number of fragments formed by disc fragmentation depends strongly \citep[e.g., fig. 8 in][]{Brucy_21_Stochastic_fragmentation} on how close the disc is to the self-regulating gravitoturbulent case \citep[also ][]{Xu_25_RHD_disc_fragmentation}. Systems far from fragmentation produce either no fragments, or rare singletons via stochastic disc fragmentation \citep[e.g.,][]{Paardekooper12-Stochastic-Fragm}; however, more massive discs that cool more rapidly may produce one to a few fragments at any given time \citep[e.g., see][in particular their fig. 2]{ZhuEtal12a}; yet more unstable discs produce $\sim$ dozens of fragments during the self-gravitating disc phase \citep[e.g.,][]{VB06,vorobyov_elbakyan2018}. For example, Fig. 1 in \cite{cha2011} shows about a dozen gas fragments born by GI in a fragmented disc\footnote{Recent 3D simulations by \cite{Brucy_21_Stochastic_fragmentation,Xu_25_RHD_disc_fragmentation} quantify fragment generation rate statistically, but note that massive embedded objects could trigger additional disc fragmentation \citep{meru2015,Cadman_22_triggered_fragm_binary}, resulting in formation of new objects that would otherwise not form.}.

\begin{table}
    \centering
    \caption{Initial orbital separation and mass of the planets and the secondary in our simulations. }
    \label{tab:pla_par}
    \begin{tabular}{lcc}
        \hline
        \textbf{Object ID} & \textbf{$R_0\,[\mathrm{au}]$} & \textbf{$m_0\,[\mj]$} \\
        \hline
        1                    & 30  & 1  \\
        2                   & 40  & 1 \\
        3                    & 60   & 1 \\
        4                    & 80   & 1  \\
        5                    & 100   & 3 \\
        6                    & 110   & 1  \\
        7                    & 120  & 1 \\
        8                    & 130  & 1   \\
        9 (Secondary) & 140  & 5, 8, 10, 15, 25, 50 \\
        \hline
    \end{tabular}
    \label{tab:inital}
\end{table}


The exact initial masses of the fragments, and their position in the disc are still poorly understood. We chose these somewhat arbitrary, but within physically plausible values discussed in \S \ref{sec:general_setup}, as given in Table \ref{tab:inital}. To recap, we  assume that planets form at tens to $\sim 100$~au distances initially, and that the secondary star forms slightly after the planets on a wider orbit. Planet orbital evolution in self-gravitating disc is quite rapid (see Fig. \ref{fig:single_tracks} below). By the time of binary formation, some of the planets would have had a chance to migrate deeper in from their initial orbits, and a minority could end up on slightly larger orbits due to stochastic interactions with spiral density arms in the disc \citep{baruteau2011}. The wide radial range of planet locations in Table \ref{tab:inital} is consistent with this scenario, and also allows us to probe the sensitivity of our results to the initial planetary orbits (cf. \S \ref{sec:HW99}).

\begin{figure}
    \includegraphics[width=0.5\textwidth]{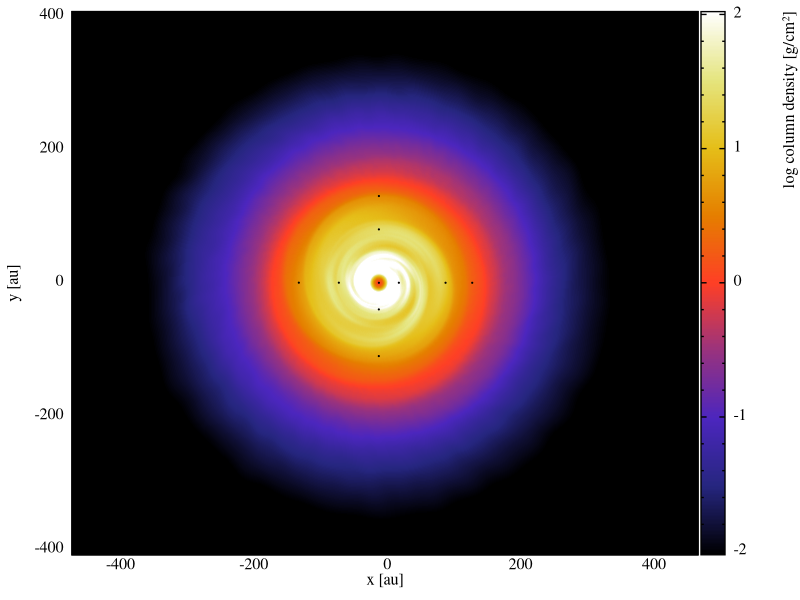}
\caption{The first snapshot of a simulation when the planets are added.}
\label{fig:initial_conditions}
\end{figure}

In terms of the initial azimuthal planet configuration, we opted to position the planets at azimuthal angles $\phi_i$ as a very simple algebraic sequence: $\phi_i = (\pi/2) (i-1)$, where $i = 1, 2, ... 9$ is the index of the planet being inserted. The initial configuration of the planets can be seen in Fig. \ref{fig:initial_conditions}. This specific choice of the initial configuration of the planets is not expected to influence our conclusions significantly in terms of statistics of the outcomes. Planet migration in gravitoturbulent discs is a rather stochastic process, with planets  experiencing inwards or outwards kicks depending on gravitational torques from the surrounding disc regions \citep[][see also \S \ref{sec:single_planet}]{machida2010,ZhuEtal12a,baruteau2011}. These planet-disc interactions are likely to dominate planet-planet interactions at least during the first few orbits of the planets. To confirm this expectation, we tested the sensitivity of our simulation results to the initial conditions by running several simulations where the planets were given a random azimuthal angle $\phi_i$, and compared these to our default initial setup. The ejection fraction did vary more in specific cases, however, it was still within the spread of ejection fractions discussed in \S \ref{sec:ejectees_statistics}.


\subsection{Object ejection criteria}\label{sec:ejection_criteria}

\begin{figure}
    \centering
    \includegraphics[width=0.5\textwidth]{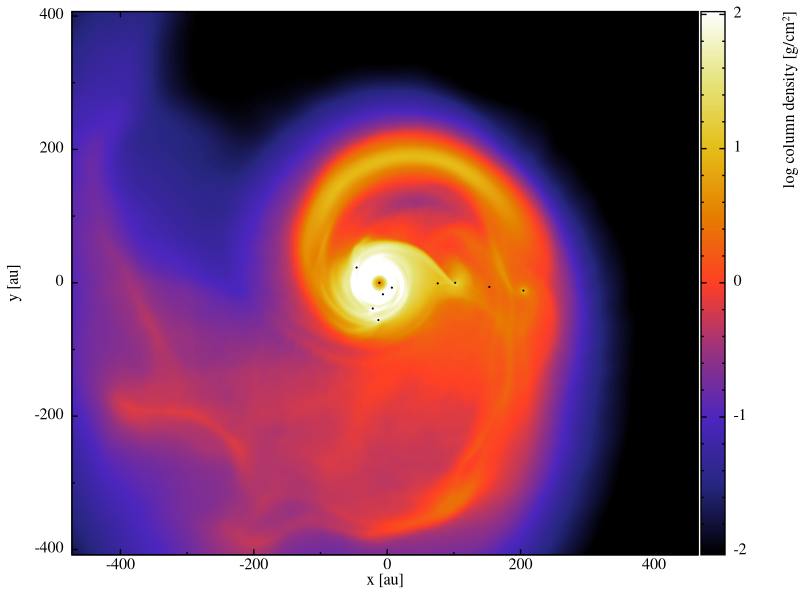}
    \caption{Same simulation as in Fig. \ref{fig:initial_conditions} but 1,400 years later, the massive object has a mass of 50 $\mj$ in this simulation.}
    \label{fig:splash2}
\end{figure}

In a planetary system without a gas disc and a massive secondary object, gravitational potential is point-like, being dominated by the central star. In such a case it is trivial to spot an ejectee: a planet with a positive total energy is unbound. For the system under consideration, however, additional care needs to be taken as part of the disc can be ejected due to interactions with the planets/secondary object (e.g., see Fig. \ref{fig:splash2}), so that our total mass distribution is often extended to radii significantly larger than the extent of our initial disc. 

To circumvent this issue, we define an ejectee as a planet that (i) reached star-planet separation of at least $10^3$~au; (ii) its radial velocity remains positive post-ejection; (iii) there is no sign of a significant radial velocity de-acceleration once the planet satisfied condition (i).

Similarly, planet eccentricity in our simulations cannot be defined based on the specific orbital energy and angular momentum using the Kepler laws for orbits around a point mass. We therefore define orbital eccentricity as a quantity averaged per orbit. We first find the sequence of apo- and peri-centres of planetary orbits, $R_{\rm a}$ and $R_{\rm p}$, respectively. Our definition of ``radius" $R$ for a planet or an SPH particle is an instantaneous distance of the former to the primary star. This distinction may be important as the star tends to move off its initial position at the coordinates' origin due to gravitational forces from the disc and the planets/secondary star. The eccentricity is then calculated by appealing to Keplerian orbital laws:
\begin{equation}
    \frac{R_{\rm p}}{R_{\rm a}} = \frac{1-e}{1+e}\;.
\end{equation}
While this definition is somewhat ad-hoc, we find it useful in visualizing orbital dynamics of our planets.

\section{Results}\label{sec:results}

\subsection{Single planet migration experiments}\label{sec:single_planet}

\begin{figure}
    \centering
    \includegraphics[width=0.5\textwidth]{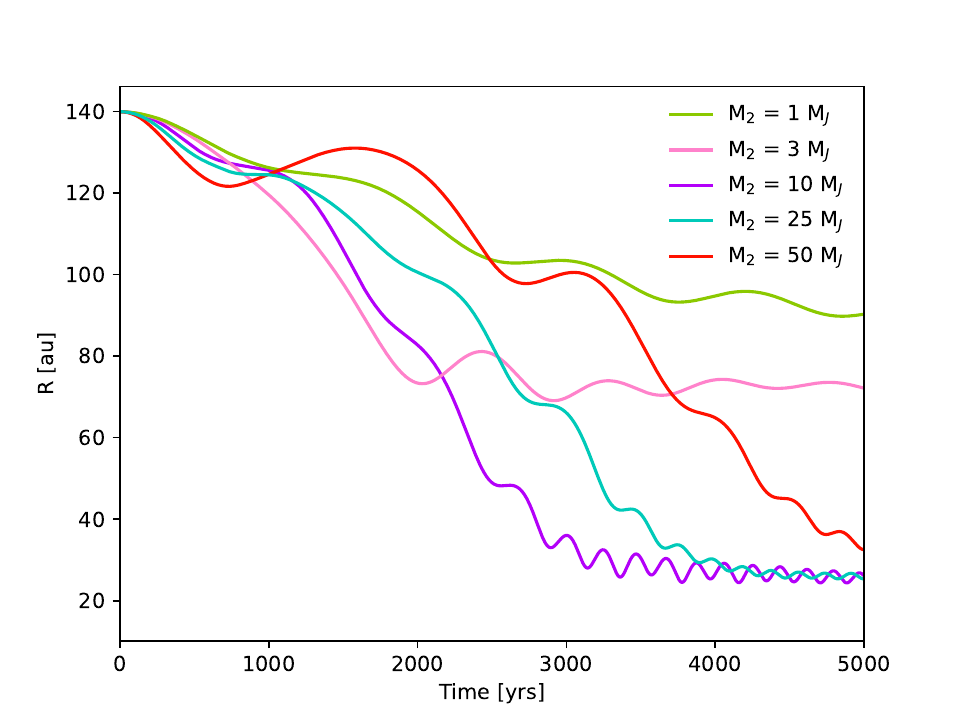}
    \caption{Migration tracks for a number of different secondary masses. The simulations all contain only one planet starting at 140 au, which is allowed to migrate for a few thousand years.}
    \label{fig:single_tracks}
\end{figure}

It is useful to consider single planet-disc interactions briefly before we present the results of the more complex multi-planet system evolution. Previous authors, e.g., \cite{baruteau2011, fletcher2019}, find that planet migration in strongly self-gravitating discs is a type-I like migration process \citep{Tanaka02,crida2006} with a superimposed chaotic component. This is due to self-gravitating discs generating strong gravitoturbulence \citep{gammie2001,rice2005} that prevents even massive planets from opening deep gaps in the disc and switching into the slower type II regime. The type I migration time scale in a laminar disc is given approximately by 
\begin{equation}
    t_{\rm mig,I} \sim \frac{3}{\Omega}\frac{M_*^2}{M_p M_{\rm D}} \Big{(} \frac{H}{R} \Big{)}^2\;,
    \label{eq:tmig}
\end{equation}
\citep[although the pre-factor in this equation does depends on the disc surface and temperature profiles, and more generally, disc physics; see, e.g.][]{Paardekooper11-typeI}. Here $M_p$, $M_*$ and $M_{\rm D}$ are the planet, stellar and disc mass respectively. This equation yields $t_{\rm mig,I} \sim$ a few thousand years for a planet of a few Jupiter masses at a distance of 100 au \citep[e.g.,][]{nayakshin2010}. The equation also predicts that planets migrate differentially, that is, more massive planets migrate inward sooner, increasing the chances of close encounters between un-equal mass planets.



Fig. \ref{fig:single_tracks} shows the migration tracks for simulations where only one planet/object was injected into the same initial disc (described in \S \ref{sec:initial}). Fig. \ref{fig:single_tracks} shows that, for the three less massive objects, the more massive of them do migrate into the inner tens of au more rapidly, as predicted by eq. \ref{eq:tmig}, than their less massive cousins. However, at higher object masses, object inertia (mass) can be comparable with the local mass of the disc, which then results in a slowing down of migration \citep{MalikEtal15}. Additionally, the migration pattern looks rather stochastic, with planets receiving inward and outward kicks that are not captured by the simple laminar linear type-I migration theory.


These results are qualitatively consistent with that of \cite{baruteau2011}, who found that injecting planets into the same self-gravitating disc on circular orbits of same initial radius but of varying azimuthal angle, $\phi$, led to  different planet migration tracks, with some of the planets even migrating outward first, before setting onto an inward migration track. Fig. \ref{fig:single_tracks} suggests that planet-disc interactions in self-gravitating discs may lead to copious close planet-secondary scatterings if their orbits cross due to the different, and rather chaotic migration patterns they experience. We also observe that the first $\sim 10^4$ years in the evolution of our planet-disc system may be the most interaction-rich epoch.


\subsection{Examples of close planet-secondary interactions}\label{sec:examples_of_interactions}

\begin{figure}
    \centering
    \includegraphics[width=0.5\textwidth]{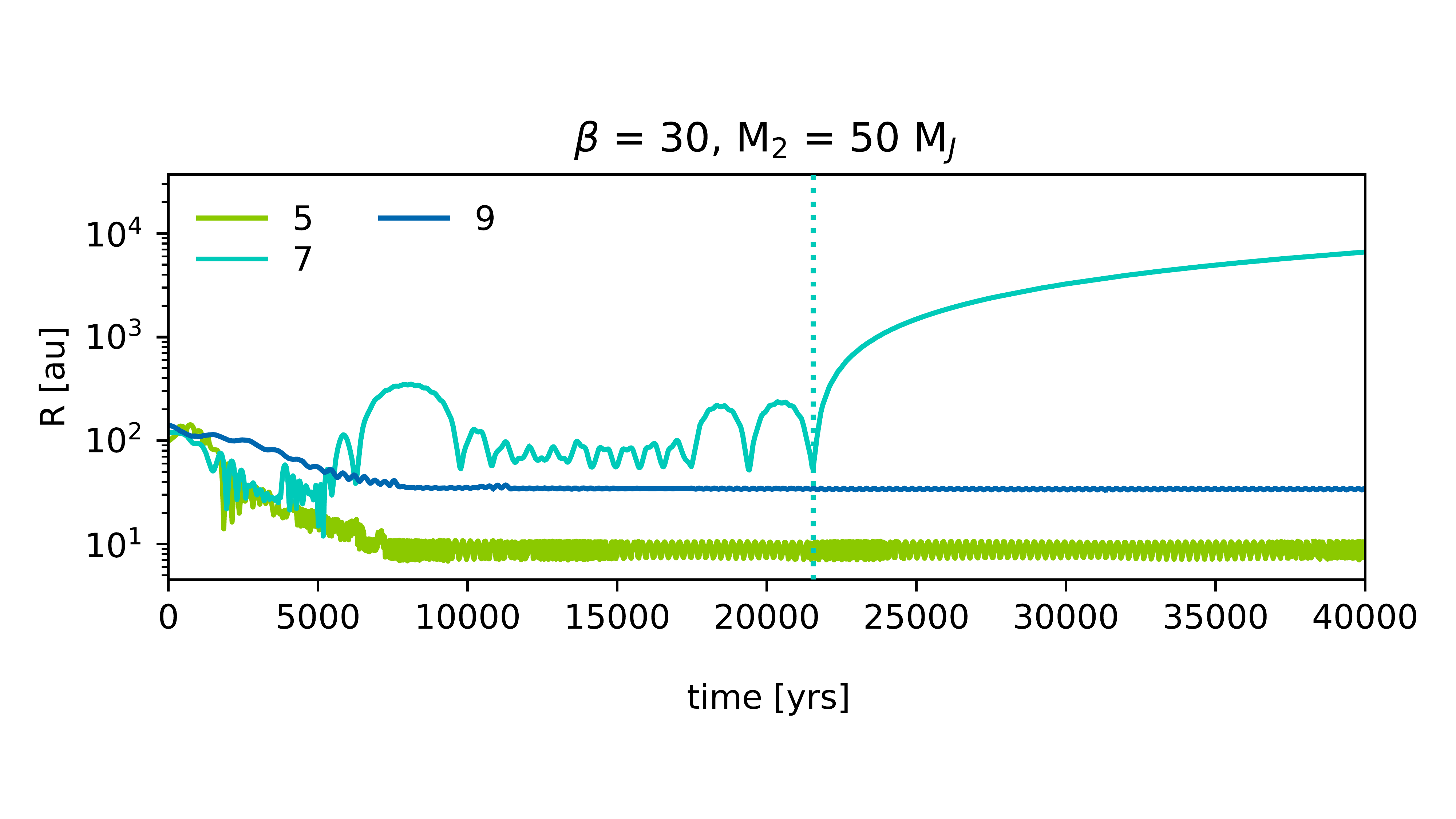}
    \caption{The radial positions with respect to the star for a subset of two planets and the secondary object in a simulation with $\beta$ = 30, $M_2$ = 50 $\mj$. The secondary has initial mass of 50 $\mj$ and is labeled ``9", while planet 7 has an initial mass of 1 $\mj$ and planet 5 starts with 3 $\mj$ (cf. Table \ref{tab:inital}). The vertical dashed line marks the time of ejection for planet 7. Purely by chance, planet 5 early interactions with the secondary send it inwards rather than outwards (cf. Fig. \ref{fig:close_encounter_59} below). It then migrates rapidly inwards, avoiding close interactions with the secondary and hence surviving as a result.}
    \label{fig:planet7}
\end{figure}

Prior to the statistical analysis of our simulation results, it is useful to consider some specific examples of planet ejections, and also of planets surviving in the planetary system.  Fig. \ref{fig:planet7} shows the radial positions (defined as the 3D distance between the star and the planet) with respect to the star and eccentricities of three objects for a simulation with $\beta$ = 30, and secondary object mass $M_2$ = 50 $\mj$. The dark blue line numbered ``9'' in the legend is the secondary. while planets ``5'' and ``7'' have masses of 3 and 1 $\mj$ respectively (as shown in table \ref{tab:inital}), which are injected into the disc at 100 and 120 au. 

\begin{figure}
    \centering
    \includegraphics[width=0.5\textwidth]{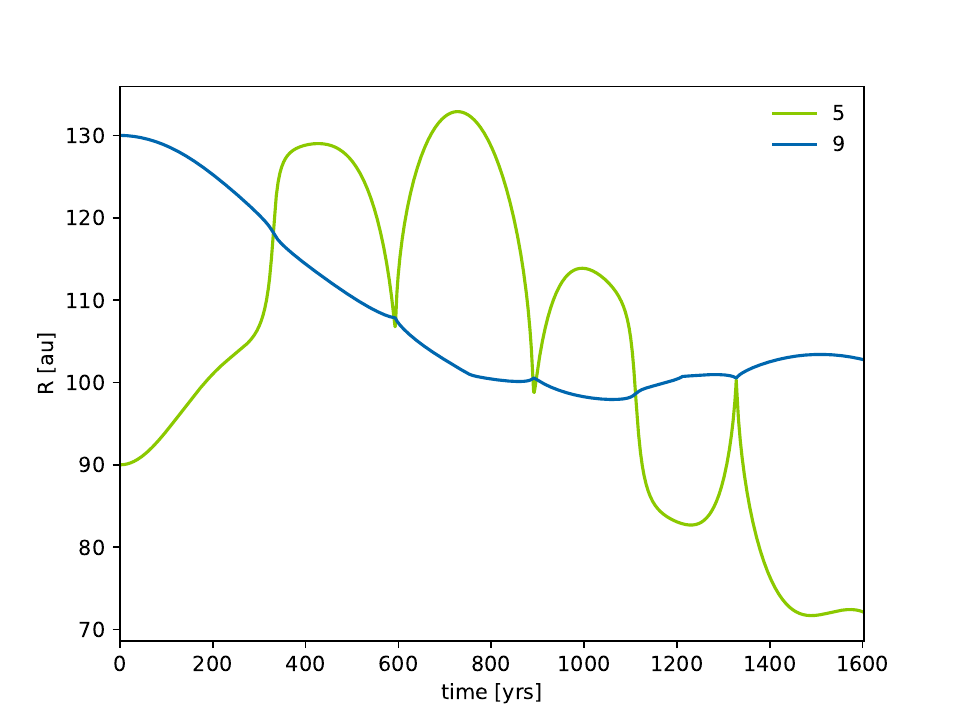}
    \caption{The radial positions of the secondary (planet 9; dark blue, $M_2$ = 50 $\mj$) and 5 (green, $M_{\rm p}$ = 3 $\mj$). An initial close encounter leads to subsequent encounters that end in planet 5 migrating inwards and escaping ejection.}
    \label{fig:close_encounter_59}
\end{figure}

We see that planet 7 first has a close encounter with the more massive planet 5, at a time just smaller than 4,000 years, which lifts its orbit to a higher semi-major axis. It then interacts with the secondary twice between 5,000--6,000 years. This dramatically alters the orbit, nearly ejecting it out of the system. Disc interactions then damp planet eccentricity somewhat; however, after a number of close encounters with the secondary, the eccentricity of planet 7 is pumped up again, and finally it receives a velocity kick strong enough to cause an ejection to infinity at time $\approx 2.2\times 10^4$ years.



Planet 5 also encounters the massive secondary early on, but survives in the system. Fig. \ref{fig:close_encounter_59} zooms in on the early time evolution of planet 5. The interactions with the secondary  lift the semi-major axis of the planet above that of the secondary for some time. However, in this case, later interactions with the secondary remove energy and angular momentum from planet 5; it receives an inward kick rather than an outward one, and eventually the  planet migrates much closer to the primary, deep inside of the secondary's orbit. 
As a result, planet 5 does not get ejected in this simulation. Qualitatively, we conclude from Fig. \ref{fig:close_encounter_59} that planets whose orbits are larger than that of the secondary at the point when its migration stalls are more likely to be ejected than those whose orbits are inside the secondary orbit. 

This evolution appears to contradict equation \ref{eq:tmig}, which predicts that the secondary should migrate inward faster than the planet, but  could be understood as following. In massive self-gravitating discs, planets migrate inward in the fast type-I like regime as long as they are unable to open deep gaps in the disc \citep{baruteau2011} quickly enough. Simulations show that giant planets usually undergo type-I migration whereas brown dwarf mass objects are more likely to open deep gaps in the discs and then ``park" on slowly evolving orbits \citep[e.g.,][]{MalikEtal15,nayakshin2017,fletcher2019}. This is consistent with Fig. \ref{fig:planet7}, where the secondary object eventually stops its rapid migration. A planet that is outside of the orbit of the secondary will continue to migrate in and therefore keep coming into close contact with the secondary. Despite a significant difference in physical scales, this situation is similar to the planet ejection scenario in a cicrumbinary disc considered by \cite{Coleman-24-FFP-simulations}. On the other hand, a planet that is inside the orbit of the secondary will continue to migrate in, often without opening a gap in the disc, hence avoiding further close interactions with the secondary object.

\begin{figure*}
    \centering
    \hspace{-60pt}
    \includegraphics[width=0.4\textwidth]{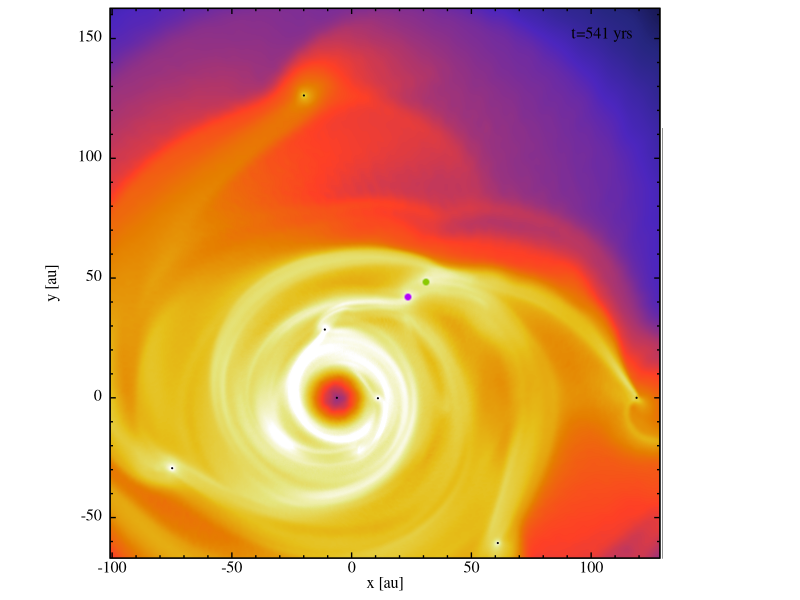}
    \hspace{-65pt}
    \includegraphics[width=0.4\textwidth]{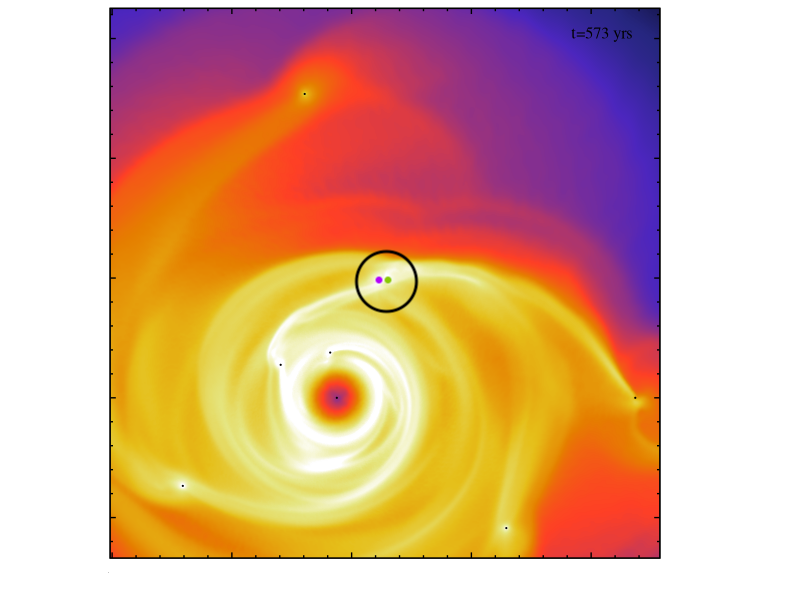}
    \hspace{-65pt}
    \includegraphics[width=0.4\textwidth]{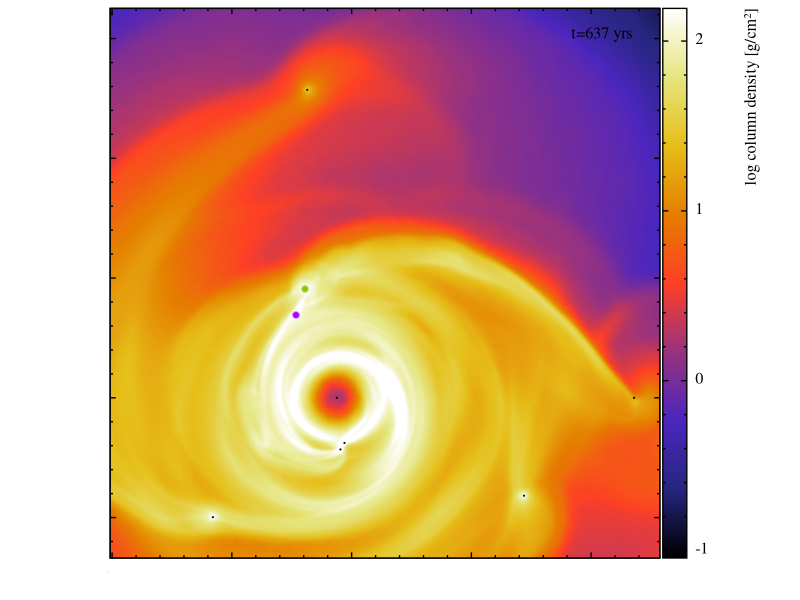}
    \hspace{-65pt}
    \caption{Three snapshots of the gas surface density profile for a simulation with $\beta$ = 20, $M_{\rm 2}$ = 25 $\mj$, at times $t=540$, 590, 640 years, from left to right panels, respectively. The larger, colored dots correspond to planets 3 (purple, initial mass, $m_i$ = 1 $\mj$) and 5 (green, $m_i$ = 3 $\mj$) in Fig. \ref{fig:beta20_m225mj}. Both of the planets are embedded into spiral density arms which appear to help to bring the two planets close together in the middle panel. Such gas-assisted close interactions alter dynamics of the planets from the pure N-body interactions regime.}
    \label{fig:spirals}
\end{figure*}

We also found interesting examples in which planet-driven spirals in the disc play a  role in enabling close planet-planet interactions. This can be seen in Fig. \ref{fig:spirals}, where two planets (purple and green dots), initially embedded into two separate gas spirals, come to interact closely in the middle panel. The planet-planet interaction at this point seems to be abetted by the two spirals merging into one, which traps the planets in a deeper potential well for a while. The planet marked with green colour is higher in mass (3 $\mj$ compared to 1 $\mj$) than the planet marked with the purple symbol.  The planets do come apart at later times.



\begin{figure}
    \centering
    \includegraphics[width=0.5\textwidth]{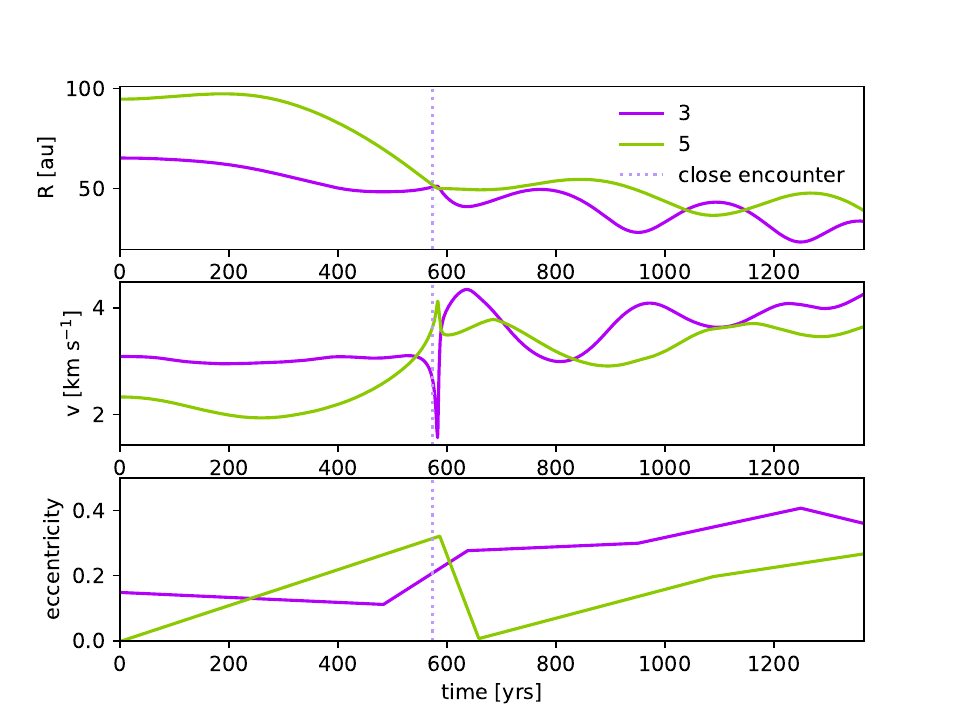}
    \caption{The planet-star distance, velocities and eccentricities of planets 3 and 5 before during and after their close encounter, which is marked by the dotted line.}
    \label{fig:close_encounter_35}
\end{figure}

Fig. \ref{fig:close_encounter_35} shows the radial positions, total velocities and eccentricities of the two planets before and after their close encounter. Before the close encounter both planets are migrating inwards, with planet 5 migrating at a faster rate due to its larger mass and greater distance from the central star. The close encounter alters the smooth nature of the initial migration, causing oscillations in the radial positions and velocities, and an increase in the eccentricity of planet 3. Migration can also occur outwards, as seen in figures \ref{fig:single_tracks} and \ref{fig:close_encounter_59}. Migration in gravitoturbulent discs is not always a uniform process, as planets may experience stochastic inward or outward kicks from spiral density arms of the disc \citep[][]{baruteau2011, boss2023}.

\subsection{Planet ejection as a function of mass of the secondary }\label{sec:ejections}

As mentioned in \S \ref{sec:add_planets} the secondary mass in our simulations ranged from (no secondary), to masses of $M_2 = $ 5, 8, 10, 15, 25 and 50 $\mj$. For each of the $M_2$ values, we have performed three simulations with identical initial conditions, while for $M_{\rm 2} =$ 0 and 5 $\mj$ two simulations were done.

\begin{figure}
    \centering
\includegraphics[width=0.5\textwidth]{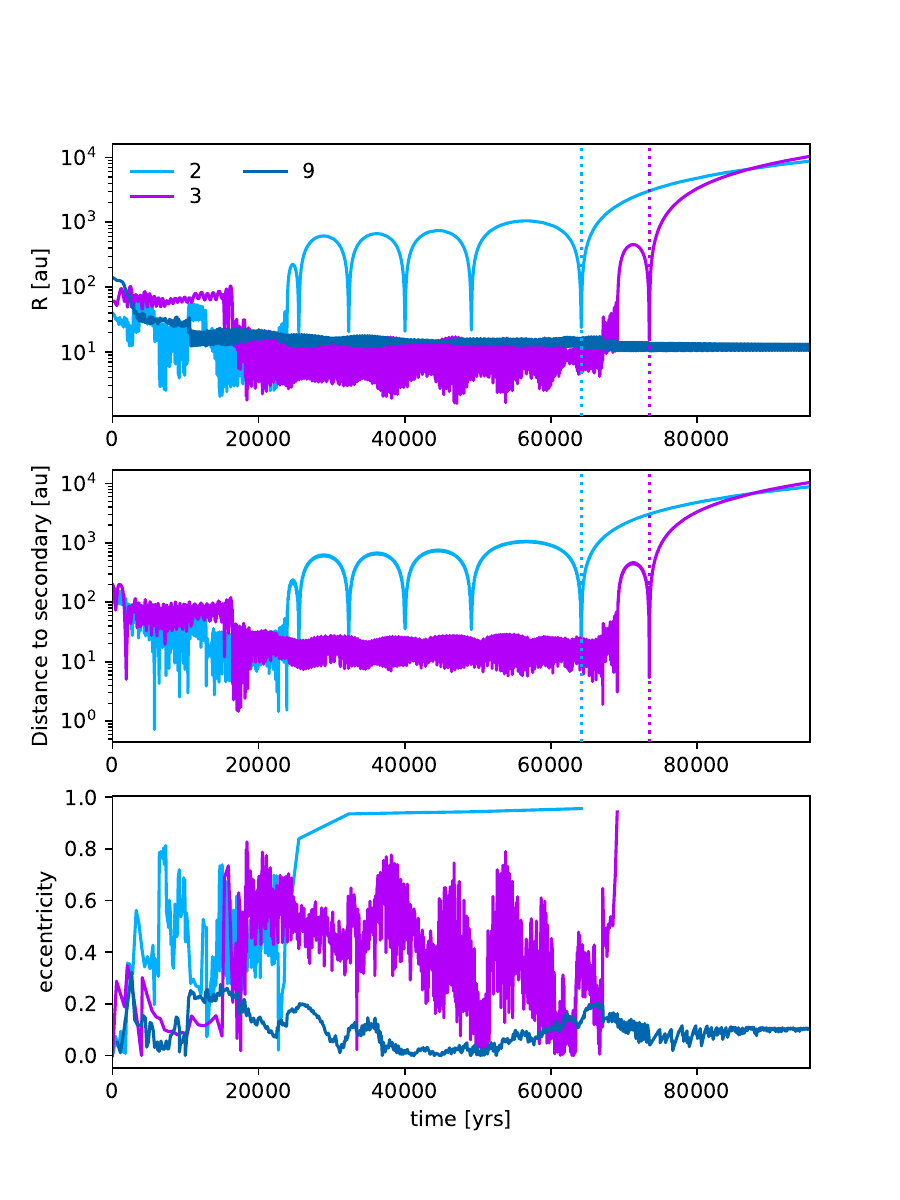}
    \caption{Time evolution of the planet radial distance from the star (top panel), the planet-secondary distance (middle panel),  and eccentricities (bottom panel). Only the planets that get ejected from the system are plotted. The dotted lines correspond to the time of planet ejections. The dark blue line (``planet 9") is the secondary of initial mass $M_2$ = 10 $\mj$. }
    \label{fig:beta20_m210mj}
\end{figure}

\begin{figure}
    \centering
\includegraphics[width=0.5\textwidth]{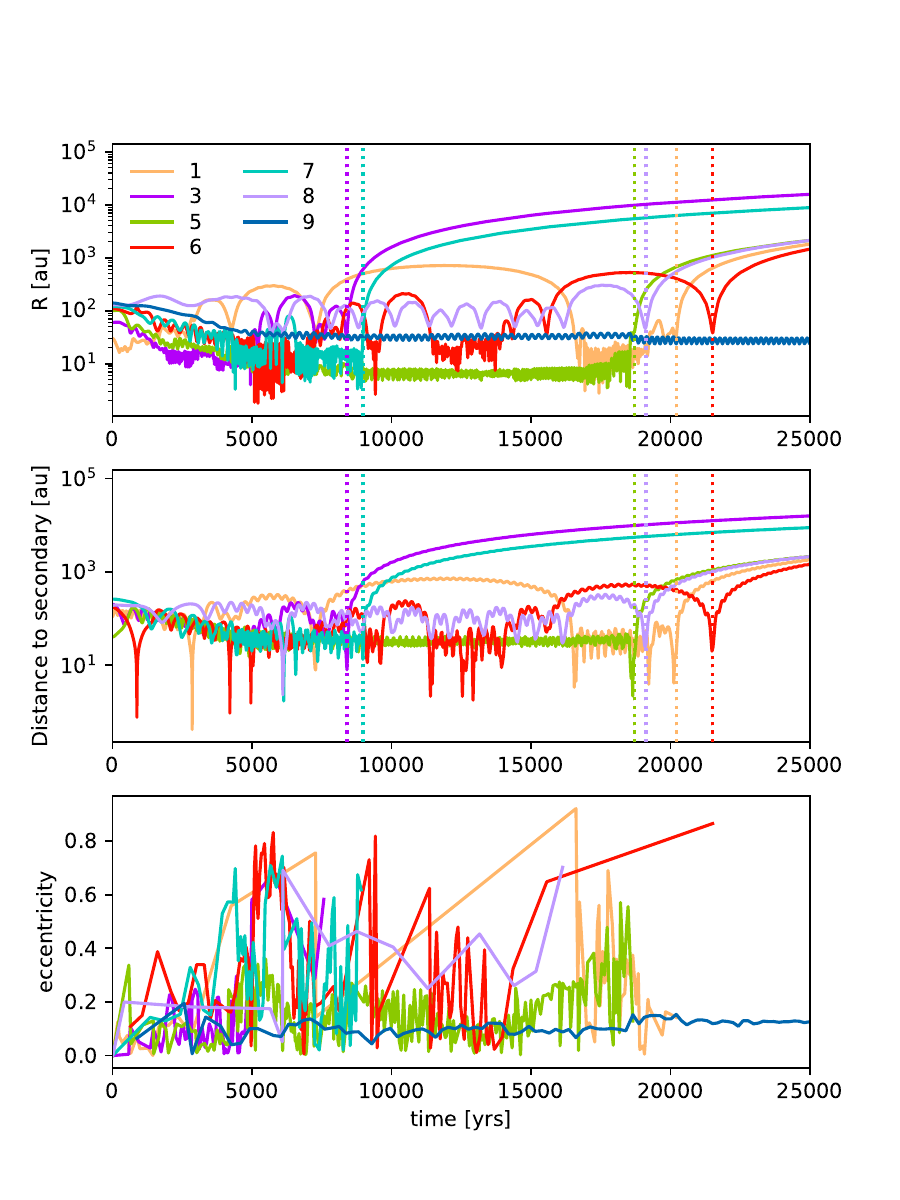}
    \caption{Same as Fig. \ref{fig:beta20_m210mj} but for $M_2$ = 25 $\mj$. Note the difference in the time axis, it is cropped here to cover the epoch of planet ejections better.}
    \label{fig:beta20_m225mj}
\end{figure}

\begin{figure}
    \centering
\includegraphics[width=0.5\textwidth]{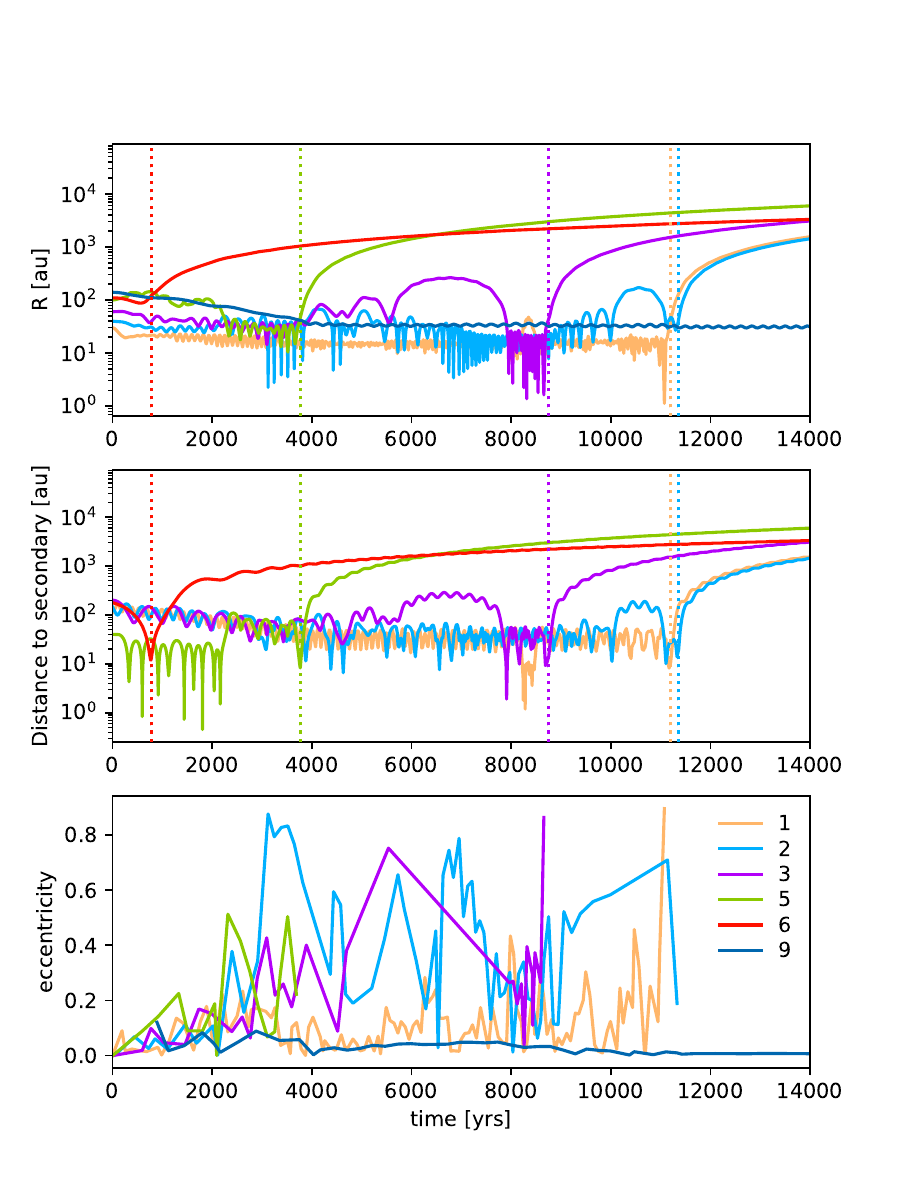}
    \caption{Same as Fig. \ref{fig:beta20_m225mj} but for $M_2$ = 50 $\mj$.}
    \label{fig:beta30_m250mj}
\end{figure}

Figures \ref{fig:beta20_m210mj}, \ref{fig:beta20_m225mj}, \ref{fig:beta30_m250mj} show the radial positions, planet-secondary distance, and planet eccentricities versus time for the ejected planets only. We do not include the planets that survived on bound orbits for figure clarity. in selected simulations for three different values of $M_2=$ 10, 25 and $50 \mj$. The planet ejection times are shown as vertical dotted lines. In \S \ref{sec:toy} we provide an approximate analytical theory for the ejection process. The ejections are a stochastic process that require (relatively) close planet-secondary interactions. These interactions provide velocity kicks to the planets. Depending on the exact nature of the planet-secondary interaction, the kick can be in the same direction as the planet velocity before the interaction, or it can be in the opposite one. The interactions can hence either increase or decrease planet's energy. When these changes are small enough to leave the planet bound, the planet-secondary interactions enforce a random walk for the planet in velocity space. On the other hand, when the velocity increases sufficiently strongly in one of the planet-secondary interactions, the planet becomes unbound, and the random walk is terminated with the planet leaving to infinity. This ejection process shares similarities with the Oort cloud formation due to close interactions of comets with Jupiter (see \verb|https://en.wikipedia.org/wiki/Oort_cloud|). This random walk can be described as one with a sharp outer edge (e.g., a particle random walk on a desk surface): while the object walks away from the boundary, the walk continues, but once it jumps beyond it, there is no coming back. For planets in the binary system, the velocity kicks are much larger than those compared to the comet-Jupiter interactions, so the random walk sequence is shorter and results in an ejection of the planets very often.

Comparing Figs. \ref{fig:beta20_m210mj}--\ref{fig:beta30_m250mj}, we observe that the higher the secondary mass, the sooner ejections occur. This can be explained by the random walk argument just discussed. The more massive is the secondary, the larger is each velocity kick (\S \ref{sec:toy}), and hence fewer planet-secondary interactions are needed for an ejection. 

The middle panels of Figs. \ref{fig:beta20_m210mj}--\ref{fig:beta30_m250mj} show that the planet-secondary minimum separation is not necessarily the defining characteristic of whether an interaction ejects the planet or not\footnote{We note in passing that the frequency of our planet and secondary data output is sufficiently high to resolve planet-secondary interactions down to the approach distance of $\lesssim 0.2$~au.}. We surmise from the figures that often an interaction that ejects the planet has a much larger minimum planet-secondary distance than previous interactions. This is due to the fact that the other variables in the interaction -- the angles between the two velocities, the exact timing of the planet-secondary interaction, and the planet's orbit before the interaction -- are also important. For example, planet 6 in Fig. \ref{fig:beta20_m225mj} (red curve) is already on a very wide orbit, with apocentre distance nearly as large as 1000 au, when it has its last interaction with the secondary. The interaction is a rather distant one (tens of au), but it is sufficient to nudge planet 6 to become completely unbound.

\begin{figure}
    \centering
    \includegraphics[width=0.5\textwidth]{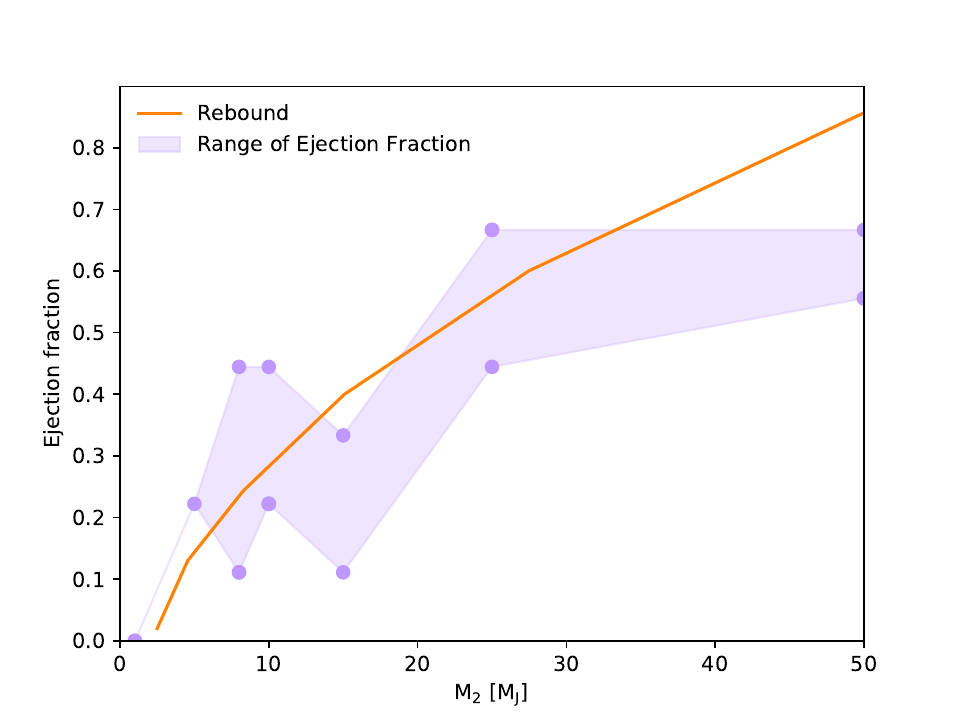}
    \caption{Purple colours: the maximum and minimum ejection fractions versus mass of the secondary object in our simulations. The higher the secondary mass, the higher is the planet ejection fraction. The orange curve shows N-body experiments of the toy model presented in Appendix B.  See \S \ref{sec:ejectees_statistics} for discussion.}
    \label{fig:ejection_frac_mass}
\end{figure}

Fig. \ref{fig:ejection_frac_mass} shows the fraction of ejected planets for all of the simulations as a function of $M_2$. The ejection fraction is defined as the ratio of the number of ejected planets to the total number of objects (excluding the central star) in the system.  For each of the values of $M_2$, for which more than one simulation was preformed, we usually have more than one value of the corresponding ejection fraction. We therefore define the maximum and the minimum of the ejection fraction obtained for a given $M_2$. In Fig. \ref{fig:ejection_frac_mass} the shaded region is the region between the minimum and the maximum ejection fractions. For $M_2=5 \mj$, the simulations had the same ejection fraction, and therefore appear as just one point on the plot. Two of the simulations in Fig. \ref{fig:ejection_frac_mass} did not have a secondary (the $M_2= 0$ point in Fig. \ref{fig:ejection_frac_mass}), and they both did not produce any ejections during the 95,000 years of the simulation. 

We observe that generally, the larger the mass of the secondary, the larger is  the ejection fraction of planets. There is a significant spread in the ejection fraction, which is related to the stochastic nature of planet-disc and planet-secondary interactions in our simulations. 
The trend of increasing ejection fraction with $M_2$ is expected, of course. The typical distance of close approach required for an ejection, derived in \S \ref{sec:analytic} (eq. \ref{rmin_scat}) increases with $M_2$ linearly. This implies that more massive secondaries find it easier to eject planets. In finer detail, Fig. \ref{fig:ejection_frac_mass} shows a local minimum in the ejection fraction at $M_2 = 15\mj$. This is explainable by the dependence of the secondary object migration history on its mass. Fig. \ref{fig:single_tracks} shows that objects of intermediate mass, $M_2 = (10-25)\mj$ in the figure, migrate inward the fastest. This imply that they have less of a chance to have a sufficiently close encounter to eject a planet.

The brown line in Fig. \ref{fig:ejection_frac_mass} presents the ejection fraction for an N-body model of one planet per disc described in Appendix B. In the model, the initial semi-major axis of the planet is 70 au, and the secondary object migrates according to an exponential friction law. The model is too simple to be expected to fit our hydrodynamical simulations. However, it is orders of magnitude faster and allows a rapid exploration of the parameter space. For each point in Fig. \ref{fig:ejection_frac_mass} we performed 300 experiments. We observe a good qualitative match, supporting robustness of our main results. The reader can download the Python code and repeat or modify the experiments.

\subsection{Statistical properties of the ejected planets}\label{sec:ejectees_statistics}

\begin{figure}
    \centering
\includegraphics[width=0.5\textwidth]{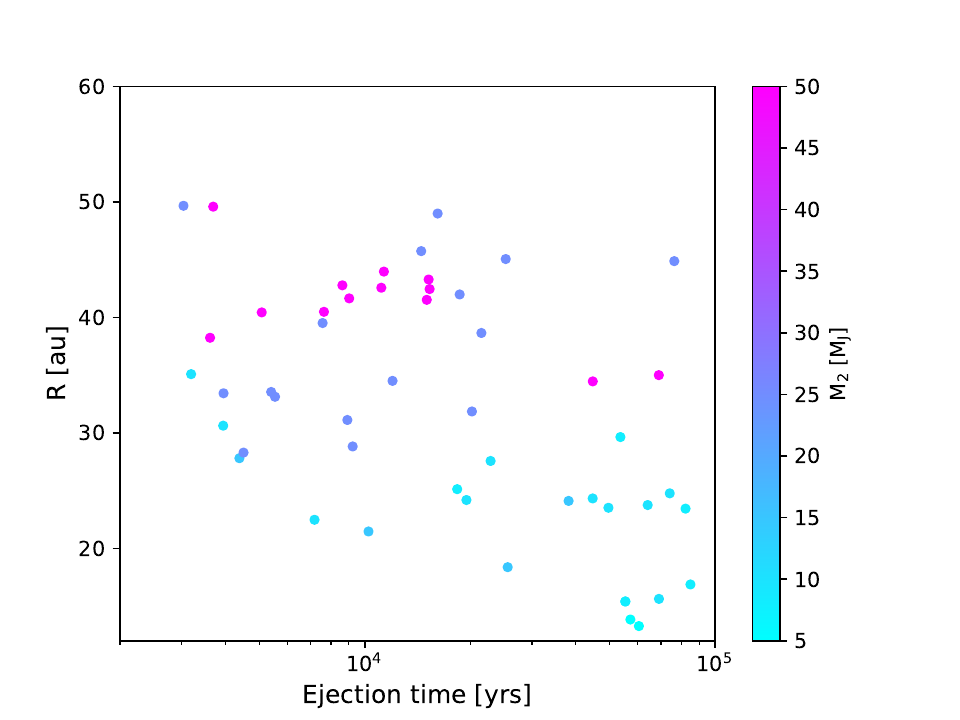}
    \caption{Radius versus time of the last planet-secondary interaction for all of the planets ejected in the simulations. Two planets ejected from much larger radii at the beginning of the simulations are omitted from the figure to better focus on the majority of the population. }
    \label{fig:time_eject}
\end{figure}

Fig. \ref{fig:time_eject} shows the kick location, $R$, defined as the distance to the central star at the time of the last planet-secondary interaction, versus ejection time for all of the ejected planets across all of our simulations. The colour bar in Fig. \ref{fig:time_eject} differentiates the simulations by the secondary mass, $M_2$. We observe that ejections at early times typically occur further out in the disc; this is simply because the secondary starts far out, at $R=140$~au at $t=0$ in our simulations, and takes some time to migrate to the inner disc. Additionally, there is also a tendency for lower mass secondaries to eject planets at later times and from closer in. This trend was already noted in Figs. \ref{fig:beta20_m210mj}--\ref{fig:beta30_m250mj}, and is likely due to the fact that lower mass secondaries require a greater number of close encounters before they are able to eject a planet. Furthermore, as we shall see later (Fig. \ref{fig:m2_a_mass}), lower mass secondaries are able to migrate closer to the central star (because they open gaps in the disc at smaller radii), whereas the more massive ones stall further out. Close planet interactions with more massive secondaries hence usually occur at larger radii simply because this is where the more massive secondary objects are.

\begin{figure}
    \centering
\includegraphics[width=0.5\textwidth]{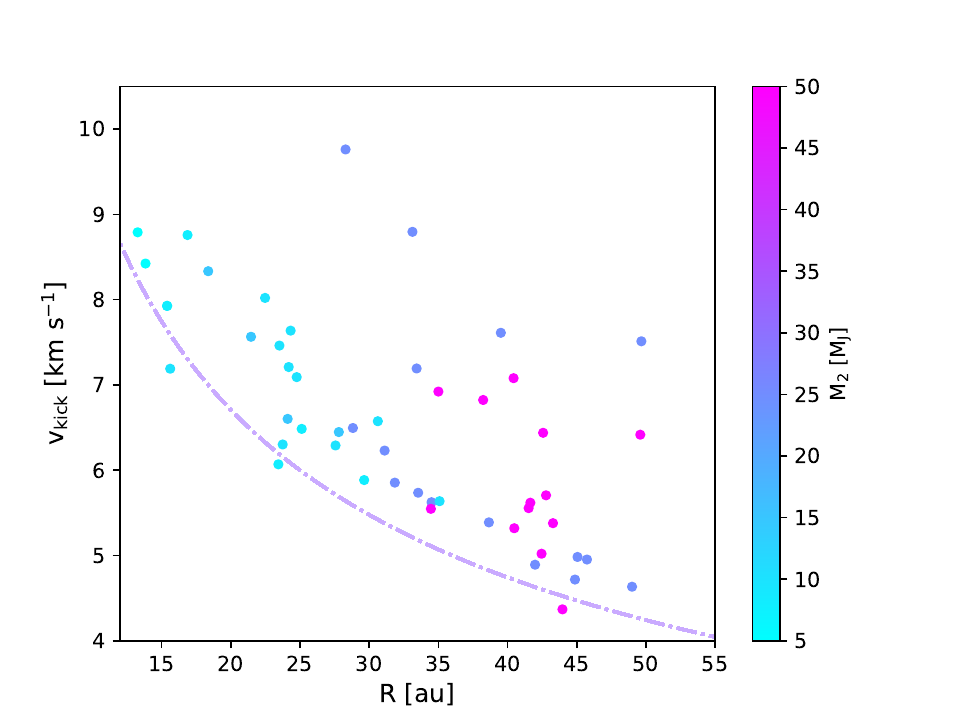}
    \caption{The kick velocity vs. kick location for ejected planets. The purple dot-dashed line represents the escape velocity, $v_{\rm esc} = \sqrt{2GM_*/R}$. We observe that the kick velocity is almost always above $v_{\rm esc}$ by just tens of \%, except for secondaries of the highest masses.}
    \label{fig:vkick}
\end{figure}

In Fig. \ref{fig:vkick} we analyse the dependence of the kick velocity, $v_{\rm kick}$, on the kick location. This velocity is the defined as the velocity of the planet at the time of ejection -- i.e., the velocity during the interaction with the secondary that unbinds the planet.
The purple dot-dashed line is the escape velocity from the primary star potential well, $v_{\rm esc} = \sqrt{2GM_*/R}$.  We can see that typically, planets are ejected with the kick velocity just barely larger than absolutely necessary for them to become unbound. We see this trend more easily for the simulations with lower secondary masses, suggesting that higher mass secondaries can impart larger kicks.

\begin{figure}
    \centering
\includegraphics[width=0.5\textwidth]{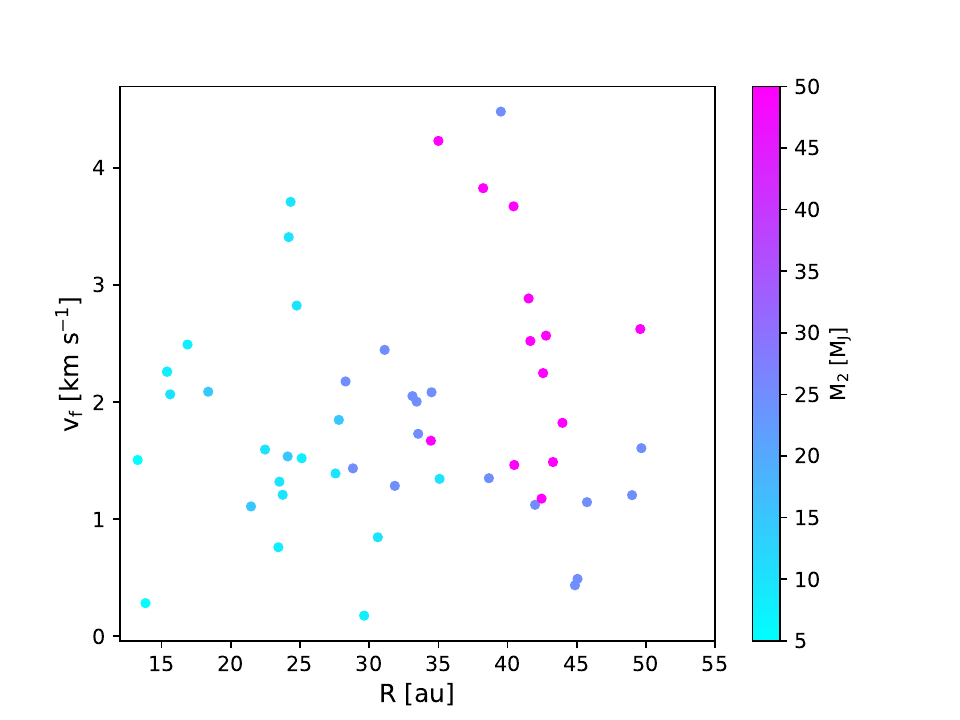}
    \caption{The final velocity vs. the kick location (defined as the location of the last close planet-secondary scattering) for ejected planets. See \S \ref{sec:ejectees_statistics} for detail.}
    \label{fig:vf_slice}
\end{figure}

\begin{figure}
    \centering
\includegraphics[width=0.5\textwidth]{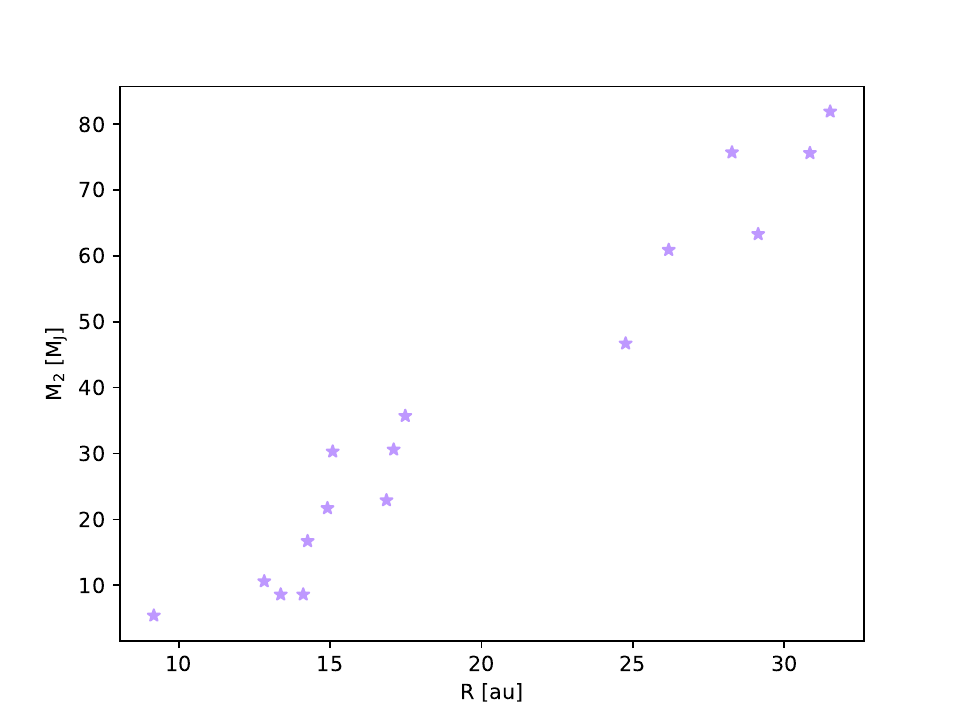}
    \caption{The final semi-major axis versus final mass of the secondary for all of our simulations. More massive secondaries open deep gaps in the disc at larger distances, and tend to be stranded on wider orbits. See \S \ref{sec:ejectees_statistics} for detail.}
    \label{fig:m2_a_mass}
\end{figure}

Fig. \ref{fig:vf_slice} presents the dependence of the final velocities of the ejected planets, $v_{\rm f}$, on the kick location, $R$. This velocity is the ejected planet velocity at the end of the simulation, defined with respect to the primary star's velocity. Fig. \ref{fig:vf_slice} shows that, unlike $v_{\rm kick}$,  $v_{\rm f}$ is not a decreasing function of the kick radius, and that the planets with the highest kick velocities in Fig. \ref{fig:vkick} are not necessarily the planets with the highest final velocity in Fig. \ref{fig:vf_slice}. For example, one of the planets in a simulation with $M_{\rm 2} = 5 \mj$ has the last close interaction with the secondary at $R = 14$ au, and receives the kick velocity of 8.4  km s$^{-1}$. The final velocity of this planet is only 0.28 km s$^{-1}$. The difference between these two velocities is obviously due to the deep potential well that the planet needs to climb out of from $R=14$~au.

It can be seen from Fig. \ref{fig:vf_slice} that the final velocities of planets in our simulations are quite low, with an average of (1.9 $\pm$ 1.0) km s$^{-1}$. These values of post-ejection velocity are in good agreement with the velocities of ejected clumps in \cite{Stamatellos_07_BDs,Basu_Vorobyov_12_BDs}. Furthermore, these velocities are in good agreement with velocity dispersion of stars in young star-forming clusters \citep[][]{Basu_Vorobyov_12_BDs}. Such velocities are much lower than the average velocity obtained by \cite{Coleman-24-FFP-simulations} in their simulations (see \S \ref{sec:velocities}). The difference in the velocities arises from the difference in the initial conditions. \cite{Coleman-24-FFP-simulations,Coleman-24-FFPs} explore a CA planet formation pathway, where planets are typically ejected from distances of order $\sim 1$~au. Since their planets escape from a much deeper potential well, they require much larger kicks to become unbound. This difference in final velocity of ejected planet is key to disentangling the two frameworks for JFFP formation via observations of young star formaing clusters (\S \ref{sec:velocities}).


Fig. \ref{fig:m2_a_mass} shows the final semi-major axis versus  final mass of the secondary for all of the simulations. Clearly, more massive secondaries end up further out in the disc. At the first glance, this contradicts equation \ref{eq:tmig}, which shows that  more massive objects migrate inwards more rapidly. However, equation \ref{eq:tmig} is only valid for relatively low mass objects (in the context of self-gravitating discs, those in the planetary mass regime). Higher mass objects open deep gaps in discs and start migrating in type II regime \citep{crida2006}, which is much slower \citep[e.g,][]{MalikEtal15}. The gap opening mass is an increasing function of radius \citep[e.g., see ][]{nayakshin2017}, and therefore more massive secondaries finish at wider orbits in our simulations. It is worth noting that in astrophysical discs this result may be weakened substantially by a continuous infall of matter from a collapsing envelope that we do not include in our simulations. The circularisation radius of incoming gas is likely to vary from system to system, and with time in the same system \citep[e.g.,][]{VB10,VB15}, and so we expect binaries to form with a range of separations whatever $M_{\rm s}/M_*$ is.

\subsection{The relation of our simulations to N-body orbital stability experiments}\label{sec:HW99}

\cite{Holman_Wiegert_99} presented  N-body simulations of planet orbital stability in a disc-less binary system. Integrating planetary orbits for $10^4$ binary orbital times, they found a range of planetary orbits that were stable to ejection from the system via close interactions with the stars, and provided a very useful fitting formula as a function of binary mass ratio and eccentricity. 

\begin{figure}
    \centering
\includegraphics[width=0.5\textwidth]{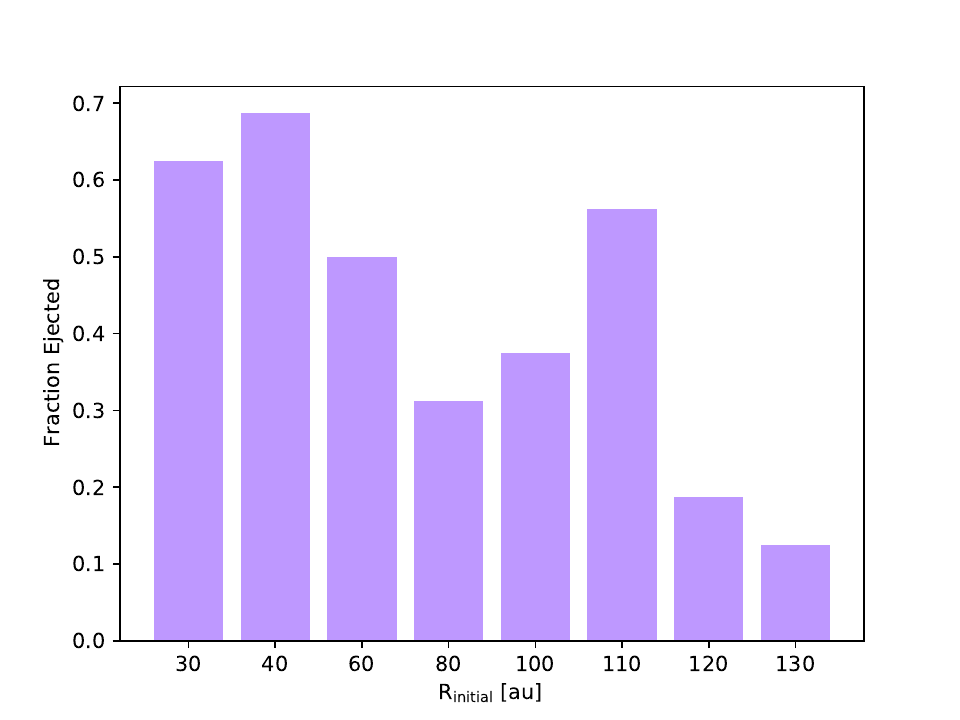}
    \caption{Histogram for showing the fraction of ejected planets per sink ID, given in terms of their initial locations in Table \ref{tab:inital}}
    \label{fig:sinkIDhist}
\end{figure}

In this paper, we start the simulations {\em before} the secondary star forms; our simulations proceed for $\sim$ hundreds of binary orbits only, and we have  extra physics -- a very massive disc that forces the orbit of the binary and the planets to evolve very rapidly. Nevertheless, the physics of planetary ejections is similar to that of \cite{Holman_Wiegert_99} -- close planet-secondary interactions -- so we may gain an insight by using their results.

The eccentricity of our secondary stars tends to be relatively low (see Figs. \ref{fig:beta20_m210mj} to \ref{fig:beta30_m250mj}), so let us set $e\approx 0$ for binary eccentricity. Eqs. (1) and (3) of \cite{Holman_Wiegert_99} then show that, for a binary with a fixed separation $a_{\rm bin}$, planet orbits with semi-major axis, $a_{\rm p}$, should be unstable to an ejection if $0.43 a_{\rm bin} \leq a_{\rm p} \leq 2.0 a_{\rm bin}$. 

Fig. \ref{fig:sinkIDhist} presents planet ejection fraction as a function of their initial orbital position, marginalized over all of the simulations with secondary mass $M_2\geq 5\mj$. Clearly, these are the planets at $a_{\rm p} = (30-60)$~au that are most liable to an ejection in our simulations. To understand this, recall that the secondary migrates inward rapidly (cf. Fig. \ref{fig:single_tracks}) from its initial position at $a_{\rm b}=140$~au to $\sim 15-30$~au (cf. Fig. \ref{fig:m2_a_mass}), where it stalls by opening a deep gap in the disc there \citep[e.g.,][]{MalikEtal15}. Since the secondary spends most  of the time during our simulations at $\sim 15-30$~au, the most unstable to ejection planets are those with the initial orbital separation of (30-60)~au. Conversely, planets 7 and 8 are the least likely to be ejected because the secondary migrates so rapidly from its initial location that it has little chance for a close interaction with the planets. The local peak at $a_{\rm p} = 110$~au is most likely due to the fact that planets at that location are affected by secondaries both during its migration phase and after it stalls. Appendix B1, the right panel of Fig. \ref{fig:Rebound}, displays a similar local peak in the ejection fraction, and reproduces Fig. \ref{fig:sinkIDhist} surprisingly well given the simplicity of the N-body model presented there.


\subsection{Surviving planets}\label{sec:survivors}

\begin{figure}
    \centering
\includegraphics[width=0.5\textwidth]{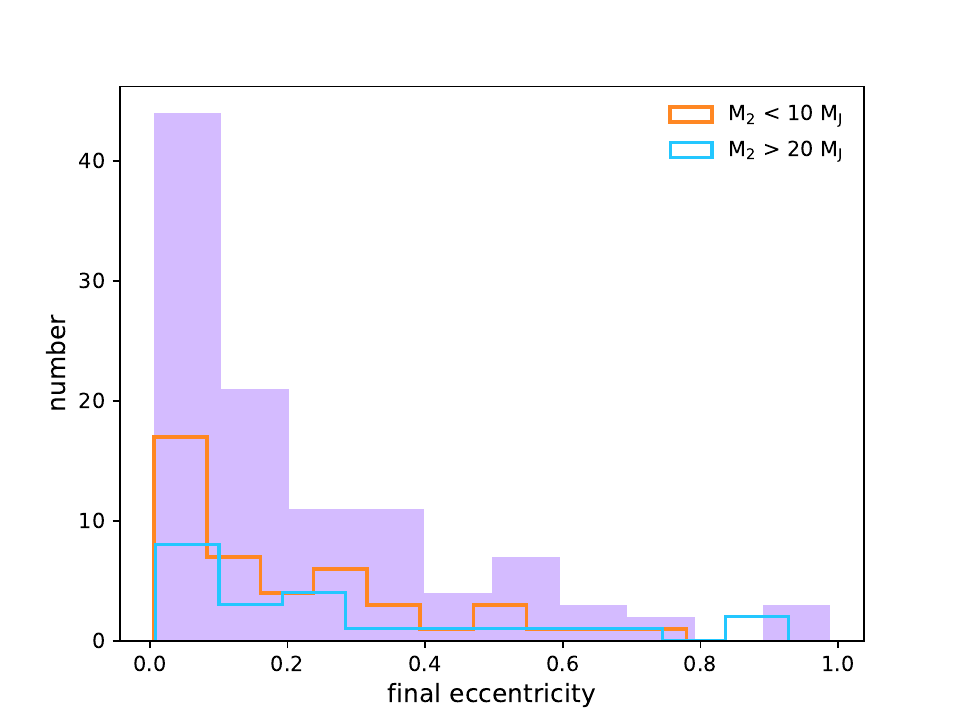}
    \caption{The final eccentricities of the surviving planets across all the simulations (purple solid bars). The orange and blue lines correspond to simulations with $M_{\rm 2} < 10 \mj$ and $M_{\rm 2} > 20 \mj$ respectively.}
    \label{fig:ecc_survivor}
\end{figure}

\begin{figure}
    \centering
\includegraphics[width=0.5\textwidth]{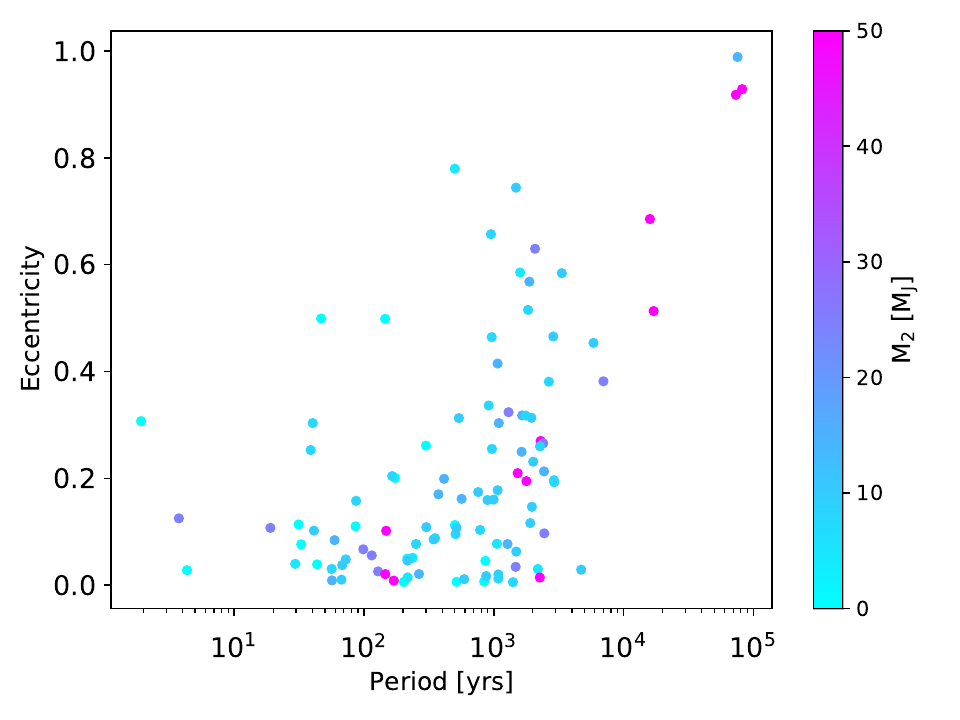}
    \caption{The final eccentricities versus period of the planet survivors, i.e., for planets still orbiting their stars at the end of the simulations. The colours depict the secondary mass (recall that two simulations had no secondary object).}
    \label{fig:period_ecc}
\end{figure}

The final eccentricity distribution of surviving planets in all of our simulations is given in Fig. \ref{fig:ecc_survivor}. The average eccentricity of the surviving planets is 0.23,  which is likely a result of planet-planet scattering and close encounters between planets as found in earlier studies \citep[e.g.,][]{weidenschilling1996,ford_rasio_2008, carrera2019, frelikh2019, marzari2024};


Fig. \ref{fig:period_ecc} shows the final eccentricity vs. orbital period for the surviving planets. We observe that the higher eccentricity planets tend to be on wider orbits, although some of the low eccentricity planets are also on far out orbits. The planets with the longest periods and largest eccentricities in Fig. \ref{fig:period_ecc} are likely to be lost in realistic astrophysical setting through other processes that we do not model here, such as flybys, and via planet-planet scattering later on \citep[][]{carrera2019, parker2024, yu2024}. Within the relatively small number statistics, our results are consistent with \cite{Bowler_20_cold_giants_ecc} for the eccentricity of cold $a = (5-100)$~au gas giant planets, who found mean eccentricity of $e\approx 0.13$.

Overall, we conclude that in systems with high secondary masses the odds of a giant planet surviving in a state bound to its star are significantly lower than being ejected. This prediction of our simulations is qualitatively consistent with observations. The incidence of gas giant planets on wide orbits \citep[tens of au;][]{ViganEtal17,Vigan20-Direct-Imaging} is a few \%, which is at least a few times lower than that of JFFPs \citep[$\sim 5-10 $\%;][]{sumi2023,gratton2025,Yee-25-FFPs-MFunction}.



\section{Analytical considerations}\label{sec:analytic}

\subsection{An order of magnitude explanation for planetary ejections}\label{sec:toy}

The physics behind planet ejections in our simulations is related to the well known gravitational sling-shot effect, also known as gravitational assist effect, that was first proposed by Friedrich Zander in 1925 as a means to give velocity kicks to interplanetary probes as they fly past one of the Solar System planets (see 
\href{https://en.wikipedia.org/wiki/Gravity_assist}{https://en.wikipedia.org/wiki/Gravity\_assist}). In that context, consider the interaction of a test-particle-like probe with a much more massive planet. In the planet's frame of reference, the probe approaches and leaves the planet with the same velocity, with a change in its direction only. In the heliocentric frame of reference, both the direction and the magnitude of the probe's velocity changes. In general, both an increase and a decrease in the probe's velocity is possible, but its possible to chose just the right angle of approach to ensure the probe's acceleration in the heliocentric frame.

In this paper, the planets play the role of the probe, and the secondary object plays the role of the much more massive object. The mechanics of the gravitational assist effect is the same, but of course the pre-interaction relative velocity, the impact parameter, and the angle of approach are all controlled by the disc torques and initial conditions of the problem, rather than the probe's engine/operator. For this reason it is the secondary, not the planets, which ``flies" through the system. Indeed, the disc interacts with the secondary much more strongly than with planets, which manifests itself in the fact that the secondary tends to migrate in faster than the planets (put another way, the migration time in the linear regime scales as $\propto 1/M_{\rm p}$, although with a large stochastic component, cf. Fig. \ref{fig:single_tracks}).

Consider now the planet-secondary interaction as a short duration scattering event in the frame of reference of the secondary of mass $M_{\rm s}\gg M_{\rm p}$. We neglect the mass of the planet, $M_{\rm p}$, in this interaction, approximating it as a test particle. Let $v_{\rm init}$, and $b$, be the planet velocity and impact parameter well before the interaction {\em in the frame of the secondary}. The scattering angle of the planet in that frame due to the interaction is
\begin{equation}
    \theta_{\rm sc} = 2\tan^{-1} \left(\frac{G M_{\rm s}}{b v_{\rm init}^2}\right)\;.
    \label{theta_s}
\end{equation}
Assume that $v_{\rm init} =\xi v_{\rm k} = \xi (G M_*/R_{\rm sec})^{1/2}$, with dimensionless parameter $0 < \xi \lesssim 1$, and $R_{\rm sec}$ is the star-secondary separation. Requiring the interaction to be sufficiently strong, so that the scattering angle is large, we let $\theta_{\rm sc} \geq \pi/2$, which then yields
\begin{equation}
    b \leq \frac{G M_{\rm s}}{v_{\rm init}^2} = \xi^{-2} \frac{M_{\rm s}}{M_*} R_{\rm sec} \;. 
    \label{b_scat}
\end{equation}
The distance of the closest approach during this interaction is $r_{\rm min} = b/(1 + \sqrt{2})$, which yields
\begin{equation}
    r_{\rm min} \approx 1.4 \hbox{ au } \left(\frac{0.3}{\xi}\right)^2 \frac{R_{\rm sec}}{50\hbox{ au}}\frac{q_s}{0.05}\;,
    \label{rmin_scat}
\end{equation}
where $q_s = (M_{\rm s}/M_*)$. This shows that the planet-secondary interaction does not need to be very close for a significant change in the planet velocity to occur. Note, however, that eqs. \ref{b_scat} and \ref{rmin_scat} depend strongly on $v_{\rm init}$ ($\xi$), which vary significantly and unpredictably in the simulations. Additionally, our derivation did not consider the case when the planet is already on a highly eccentric orbit before the last scattering event with the secondary. In such a case, only a small increase in the planet's velocity (energy) is required for it to become unbound as a result of the scattering. Indeed, in the simulations performed in this paper we find that there is a wide range of conditions under which the planet-secondary interactions occur, and that in some cases the closest approach distance ($r_{\rm min}$) exceeds 10~au (cf. Figs. \ref{fig:beta20_m210mj}--\ref{fig:beta30_m250mj}). Conversely, due to stochastic nature of the planet-secondary scattering events in the simulations, there are also interactions closer than eq. \ref{b_scat} that do not eject the planet and instead reduce the planet's energy, making its orbit more bound (e.g., Fig. \ref{fig:close_encounter_59}). In these cases the velocity kick from the planet-secondary scattering is in the opposite direction to the planet pre-collision velocity.

\subsection{Planet disruptions}\label{sec:disruptions}

In the simulations in this paper we treated the planets as point particles. Real planets have finite sizes, which in the case of planet formation by gravitational instability vary from clumps as large a few au in radius, $R_{\rm p}$, at birth, to $R_{\rm p}\sim 10^{-3}$~au ($\approx 2 \rj$) after they mature into dense Jupiter like planets \citep{Kuiper51b,Boss97}. The physical reason for this very wide range in sizes is their internal structure. At birth, clumps are synthesized from the material of a very cold ($T$ $\sim 20$~K) disc, and, in terms of composition are dominated by molecular Hydrogen \citep{cha2011,ZhuEtal12a}. They cool and contract with time, collapsing into much denser atomic or ionized Hydrogen-dominated planets when their central temperature reaches $\sim 2,000$~K \citep[e.g.,][]{Bodenheimer74,BodenheimerEtal80,HS08,Nayakshin_Review}. These diffuse clumps are often called pre-collapse planets. Fig. 2 in \cite{HumphriesEtal19} shows contraction tracks of pre-collapse planets of masses between 1 and 7 $\mj$; such planets tend to have radii from $R_{\rm p} \sim 0.2$ to a few au, depending on their mass and age.  

After collapse, the clumps reach central temperature over $10^4$~K \citep{Bodenheimer74}, and their radii are in tens of $\rj$ ($R_{\rm p}\sim 0.01$~au). The manner and time scales of this collapse depend strongly on topics that are well outside the scope of our simulations.
Depending on dust inventory of their atmospheres, mass, and insolation, these planets contract to $R_{\rm p}\sim (2-3) \rj$ in the course of next $10^4-10^6$ years \citep[e.g., see Fig. 16 in][]{Nayakshin-23-FUOR}. 

Let a planet with mass and radius, $M_{\rm p}$ and $R_{\rm p}$, respectively, swing by the secondary on a trajectory with the minimum approach distance $r_{\rm min}$. We define the ``secondary Hill sphere" of the planet with respect to the secondary during the close interaction as
\begin{equation}
    R'_{\rm Hill} = r_{\rm min} \left(\frac{M_{\rm p}}{3 M_{\rm s}}\right)^{1/3}\;.
\end{equation}
If $R_{\rm p} > R'_{\rm Hill}$, the secondary is disrupted. Using eq. \ref{rmin_scat}, we obtain for this maximum radius 
\begin{equation}
    R_{\rm p} \gtrsim 0.26 \hbox{ au } \left(\frac{0.3}{\xi}\right)^2 \frac{R_{\rm sec}}{50\hbox{ au}}\left(\frac{q_s}{0.05}\right)^{2/3} \left(\frac{M_{\rm p}}{1 \mj}\right)^{1/3}\;.
    \label{Rp_disr}
\end{equation}
Realistically, $R_{\rm p}$ needs to be even smaller due to tidal heating of the planet during the secondary star passage \citep[cf.][in the context of stars passing supermassive blackholes on close orbits]{Li_Loeb_13_tidal_heating}. Furthermore, consulting Figs. \ref{fig:beta20_m210mj}--\ref{fig:beta30_m250mj} we see that, even though last planet-secondary interaction may be a distant one, most planets experience approaches of $\sim 1$~au or less prior to that. 

Therefore, we conclude that if we consider pre-collapse planets (clumps) in our simulations, then most of them would be disrupted rather than ejected! We note that after our paper was submitted, \cite{Schib_25_dipsy_1,Schib_25_dipsy_2} presented GI population synthesis modeling that takes into account planet-planet interactions, indeed finding that many pre-collapse fragments are destroyed. At the same time, we also see that it is unlikely but not completely impossible for pre-collapse H$_2$ dominated planets to survive close interactions with secondary stars and be ejected as JFFPs. Indeed, such events were observed in hydrodynamical disc simulations by \cite{Stamatellos_07_BDs,vorobyov_basu2013}, although in these simulations the ejected fragment masses are generally in the BD rather than in the planetary mass regime, and the frequency of such events in realistic discs is unclear. Future dedicated simulations are needed to quantify the likelihood of pre-collapse planet ejections, but we note in passing that such pre-collapse FFPs are completely unique to the disc fragmentation scenario as they cannot be produced by the CA theory, where gas giant planets are always in the post-collapse like configuration, that is, are as small as $R_{\rm p}\sim 2 \rj$ \citep{PollackEtal96,MordasiniEtal12a}.


\subsection{Circum-planetary disc retention}\label{sec:disc_retention}

Consider now post-collapse planets with radii much smaller than $r_{\rm min}$. These planets are not disrupted when they swing by secondary stars, and they could potentially host circum-planetary discs (CPDs). Our simulations lack numerical resolution to model small-scale CPDs, e.g., those smaller than $R\lesssim 1$~au, as simulating discs on these scales would slow down our simulations significantly. However, based on an argument similar to that from \S \ref{sec:disruptions}, retention of CPDs by ejected planets is possible but not very likely. Prior to the ejection, the size of the disc is likely to be around $1/3$ of the Hill sphere of the planet around the primary star, so a few au typically. Planet-secondary close scatterings are likely to truncate the disc to about $R'_{\rm Hill}$ or somewhat less \citep[see, e.g., ][]{Hall_96_disc_truncation}. 

Many of our planets experience interactions with the secondary in which $r_{\rm min}$ is about 1 au or less, but some are lucky to escape with $r_{\rm min}\sim 10$~au (cf. Figs. \ref{fig:beta20_m210mj}--\ref{fig:beta30_m250mj}). In terms of $R'_{\rm Hill}$, this corresponds to the maximum CPD sizes of about a fifth of $r_{\rm min}$, i.e., $0.2-2$~au. Hence our ejected planets are not likely to carry away discs larger than $1-2$~au in size.

\section{Discussion}\label{sec:discussion}

\subsection{Summary of main results}\label{sec:summary_in_discussion} 

We have used 3D SPH simulations of a massive, self-gravitating disc with embedded planets, and a more massive secondary object, to study the process of ejection of planets out of the system. In our model, the secondary object is the seed of the secondary star that will grow into a low mass star by gas accretion. We experimented with initial secondary masses from very low mass, $M_2 = 5 \mj$, to $50 \mj$. Our main findings can be summarized as following: 
\begin{enumerate}
    \item Rapid and chaotic radial migration of planets and the secondary in self-gravitating discs leads to abundant orbital crossings. We find these crossings result in strong velocity kicks to the planets, providing an efficient mechanism (via gravitational assist effect, \S \ref{sec:toy}) for JFFP ejections in very young protoplanetary discs ($t < 10^5$ years).
    \item For secondary masses in the BD mass range and above, the ejection fraction is  large, i.e., between $\sim 30$ and 70\%. 
   \item Final velocities of the ejected planets are low, $v_{\rm f} = (1.9 \pm$ 1.0) km s$^{-1}$, as expected for ejections from tens to a hundred au distances from the parent star.
    \item Stellar binary systems with fragmenting discs are a very efficient source of JFFPs if these planets are in the dense post-collapse configuration. In contrast, if the planets are still in the clump-like (pre-collapse, molecular Hydrogen dominated) stage, then most of the clumps are likely to be disrupted by tidal interactions with the secondary, rather than being ejected (see \S \ref{sec:disruptions}).
    \item Circum-planetary discs of ejected planets are not likely to be larger than $\sim 2$~au at most (cf. \S \ref{sec:disc_retention}).
    \item The mean eccentricity of surviving planets in the simulations is 0.23. Although we do not follow up the surviving planet evolution after the gas disc is dispersed, the relatively high eccentricities of these systems likely mean that a good fraction of these could also be eventually lost.
 
\end{enumerate}


\subsection{Observational implications}\label{sec:observational_tests}

\subsubsection{Star-like vs  planet-like formation channels}\label{sec:as_stars}

Soon after the discovery of BD-mass FFPs \citep{Lucas_00_FFP_discovery,Zapatero_00_JFFP_discovery} many theorists voiced doubts that BDs could form by a direct gravitational collapse of a molecular gas cloud since this requires very high gas densities, proposing alternative scenarios for their formation \citep[e.g.,][]{Reipurt_Clarke_01_BDs,Bate_02_BDs,WZ_04_BD_formation}. However, as recently reviewed by \cite{Palau_24_BD_formation_review}, such high densities could possibly be reached due to compression in highly turbulent flows \citep[e.g.,][]{Padoan_02_BD_turb_fragm,Hennebelle_08_turb_ISM_theory,Haugbolle_18_turb_star_formation}. In a realistic star cluster formation, it is likely that multiple of these mechanisms operate \citep[e.g.,][]{Bate-19-Star-Cluster-formation}. The star-like formation process can thus be defined as fragmentation of a molecular cloud plus possible ejection or other enviromental ways of gas accretion termination. In contrast, the planet-like formation scenario assumes forming the object in a protoplanetary disc and then ejecting it.

Below we contrast our simulations with observations and also the 
CA \citep[e.g.,][]{Coleman-23-planet-in-CB-systems,Coleman-24-FFP-simulations,Coleman-24-FFPs} FFP formation channel.

\subsubsection{Bulk FFP metal abundance}\label{sec:bulkZ_discussion}

Bulk metal abundance of JFFPs may be a sensitive probe of the planet-like vs star-like formation channels. A population of FFPs formed as stars should be expected to have metallicity distribution similar to that of stars in the same cluster. On the other hand, if FFPs form as planets, then we would expect their metallicity to be elevated, whether they formed via CA or disc fragmentation. Observations show that bound gas giant planets  have metallicities larger than that of their star by a factor of a few \citep[e.g.,][]{ThorngrenEtal15}. These observations relate to planets on very short period orbits, however, it is not clear where exactly these planets formed. In the CA scenario for planet formation, accretion of planetesimals by the growing planet envelope \citep[e.g., see][and references therein]{Shibata_22_Z_enrichment_CA} may bring extra metals into the planet. Metal enrichment of disc fragmentation planets is quite efficient and occurs via pebble accretion  \citep{BoleyDurisen10,humphries2018,Vorobyov-Elbakyan-19,Baehr19-pebble-accretion}. 

\subsubsection{Disc presence}\label{sec:disc_presence_discussion}

Near-infrared observations of young nearby star forming clusters indicated early on that some FFPs carry protoplanetary-like discs \citep[e.g.,][]{Luhman_05_FFP_with_disc}. The disc-bearing objects are usually more massive than $\sim 10 \mj$, although \cite{Langeveld-24-JFFPs} show a tentative JWST detection of a disc around an FFP with mass as low as $\sim 5 \mj$. Furthermore, the observed disc fraction in young FFPs is large, $\gtrsim 40$\%, does not decrease towards masses possibly as low as $\sim 5\mj$ \citep{Scholz_23_NGC1333}, and declines with age at a rate apparently similar to that of protoplanetary discs around stars \citep[cf. Fig. 5 in][]{Seo_Scholz_25}. These observations provide strong support for a significant fraction of massive FFPs forming in a star-like fashion.

Below a FFP mass of a few $\mj$, infrared observations of young FFPs remain difficult, and presence of discs is not yet constrained \citep{Langeveld-24-JFFPs}. The two planet-forming channels for FFP formation differ significantly in their ability to create disc-bearing FFPs. If young FFPs with masses below opacity fragmentation limit \citep[$\sim 5 \mj$, ][]{Low76,Rees76} are observed to carry discs then it would be a strong argument in favor of the disc fragmentation FFP formation scenario.

In particular, in the CA planet formation framework, the most promising environment for creating an abundant population of FFPs are compact stellar binary systems, i.e., those with separations on the order of an au \citep{Coleman-24-FFPs}. Planets are then kicked out by a close interaction with the secondary star, similar to the disc fragmentation picture. Due to the much smaller scales involved, and much larger velocity kicks needed for planet ejections from close binary systems, the discs retained by CA FFPs would be minuscule, i.e., significantly smaller than 0.1 au in radius.

We argued in \S \ref{sec:disc_retention} that planets ejected from our discs may carry discs no larger than $\sim 2$~au. On the other hand, this conclusion depends sensitively on the scales from which the ejections occur. \cite{Basu_Vorobyov_12_BDs} and \cite{VB15} study disc fragmentation in 2D simulations that include protoplanetary disc formation from first principles from envelope collapse, and find that their discs can fragment on scales of hundreds of au \citep[e.g., see Fig. 1 in][]{VB15}. This is significantly larger than the scales on which we embed planets in our simulations. \cite{Basu_Vorobyov_12_BDs} find multiple fragment ejections. While they do not model second collapse of the fragments, their fragments themselves can be as large as $\sim 10$~au in radius (and some can be as massive as $0.1\msun$). It is hence potentially possible that disc fragmentation FFPs could carry protoplanetary-like discs as large as $5-10$ au\footnote{Additionally, note that internally ejected gas clumps \citep{Basu_Vorobyov_12_BDs} and stellar flybys \citep{Fu_25_FFP_from_flybys} can pick up and stretch a significant fraction of the protoplanetary disc into a filament. These filaments can then collapse into massive planets with large discs around them due to the unusually high rotational-to-gravitational energy ratio of gas in such filaments \citep{Basu_Vorobyov_12_BDs}.}.

\subsubsection{JFFP numbers per field star}\label{sec:numbers_discussion}

As we mentioned in \S \ref{sec:as_stars}, there is an uncertainty on whether JFFPs form as stars or as planets \citep{Palau_24_BD_formation_review}. One can try and differentiate between the star-like and planet-like formation channels by considering  the observed mass functions of FFPs, BDs and stars \citep[e.g.,][]{Yee-25-FFPs-MFunction}. In doing so, one singles out objects that formed as planets, as belonging to the  the steeply declining power-law of the microlensing FFPs \citep{Gould-22-microlensingFFPs,sumi2023,Mroz-23-Microlensing-planets-review}. In contrast, the objects that formed as stars are identified via extrapolation of the mass function downward in mass from the stellar end. However, this procedure is not without drawbacks. Firstly, the sub-stellar IMF remains a hotly debated topic and may vary from one cluster to another \citep{miret-roig2022,Miret-Roig23-FFP-Review,Kirkpatrick_24_IMF} or to the field \citep{Chabrier_05_IMF,Chabrier_23_IMF}. Secondly, how do we assign individual observed FFPs to either the planet-like or star-like channels in the range where the two different mass functions merge? 

Nevertheless, \cite{Yee-25-FFPs-MFunction} find that, within current observational uncertainties, there is between $0.075$ and $\sim 0.12$ JFFP per field star in the planet mass range between Saturn mass ($\approx 0.3\mj$) and $13\mj$. Most of these planets are sub $1\mj$ mass, so are rather likely to belong to the planet-like population. In the $1-13 \mj$ mass range, focusing on just the planet-like formation scenario, \cite{Yee-25-FFPs-MFunction} find $\sim 0.03$ FFP per star. 

However, there is further observational uncertainty. It is possible that some of the observed microlensing FFPs are actually bound planets on very wide orbits \citep[cf.][]{Gould-22-microlensingFFPs,sumi2023}. \cite{Yee-25-FFPs-MFunction} propose that all of the $\mu$FFPs are such, based on the rough coincidence in the numbers of $\mu$FFPs and planets bound to a host (per star). This extreme view is rather unlikely. \cite{Mroz-24-evidence-for-FFPs} reported deep Keck telescope observations of five such microlensing event fields, and found that none of them showed presence of a stellar mass host, although more and deeper observations are still needed to confirm this conclusion. Furthermore, direct imaging of nearby young clusters detects a similar abundance of FFPs \citep{miret-roig2022}, and these FFPs are not bound to a stellar host. Finally, \cite{Mroz-23-Microlensing-planets-review} notes that the available photometric data for observed $\mu$FFPs rule out putative stellar hosts within about 12 au of the planet. At separations beyond about 10 au, however, the abundance of gas giant planets decreases sharply \citep{Fulton_21_giants_occurence}.

The key point we make in this section is this: whether most of JFFPs are truly unbound objects or are bound planets on wide orbits, the disc fragmentation scenario that we propose here is far more promising to account for the surprisingly large abundance of FFPs than CA-based scenarios.

Consider first the case where most of observed JFFPs are actually bound to a host star. Most of the planets formed by CA population synthesis reside on relatively small (i.e., separation of a few au) orbits \citep[e.g.,][]{IdaLin04b,IdaLin08,AlibertEtal13,ma2016,NduguEtal19,Bern20-2,Emsenhuber_23_CA}, so these models would have to undergo a major re-evaluation to produce a similarly abundant population  at very wide separations. Furthermore, microlensing surveys report FFP abundance per field star; these are dominated by M-dwarfs, which are $\sim 4$ times more abundant than the FGK stars \citep[e.g., Table 1 in][]{Yee-25-FFPs-MFunction}. Both planetesimal-based and pebble-accretion based CA population synthesis predicts a strong dearth of gas giant planets around M-dwarf stars, because their discs contain insufficient amounts of metals and gas to make these massive planets \citep[][]{IdaLin_05_Low_Mstar,Liu_19_CA_low_Mstar,Burn_21_BernCA_lowMstar}. For example, \cite{Liu_19_CA_low_Mstar,Miguel_20_CA_Mdwarfs,Mulders_21_CA_Mdwarfs} find in their populations synthesis models no gas giants for stellar hosts with masses $M_* \leq 0.3\msun$.  To remedy this deficiency of CA models, disc fragmentation is invoked to form the gas giants observed around M-dwarf stars \citep[e.g.,][]{Schlecker_22_giants_in_Mdwarfs,Bryant_25_giant_around_Mdwarf}. Indeed, there is no fundamental obstacle to forming gas giant planets around M-dwarfs by disc fragmentation. For example, Vorobyov and collaborators often find, e.g., \cite{VB10,VB15}, that discs around growing protostars fragment early on, when the mass of the central protostar is significantly smaller than 1$\msun$. 

Let us now consider the case in which the observed JFFPs are a truly unbound population. Some of CA challenges here are the same as above (difficult to form enough gas giants around M-dwarfs, the most abundant type of star in the Galaxy). In terms of ejection mechanisms, earlier work has showed that planet-planet scattering may eject planets very efficiently \citep[e.g.,][]{weidenschilling1996,Rasio_96_FFPs,juric2008,Chatterjee_08_PP_scattering} from systems of many closely-packed planetary systems. However, such configurations must be quite rare because the observed occurrence rate of gas giant planets integrated over all separation is $\lesssim 0.25$ in FGK stars \citep{Fulton_21_giants_occurence}, and is much lower in M-dwarf stars. For these reasons, the planet-planet scattering channel is no longer considered viable to yield enough JFFPs in the CA scenario \citep{veras2012,ma2016,Coleman-24-FFP-simulations}.

\cite{PierensNelson08} showed that planets forming in circumbinary discs in binary systems undergo close scatterings with the secondary, and are ejected, provided they migrate close enough to the inner edge of the circumbinary disc. \cite{Coleman-23-planet-in-CB-systems,Coleman-24-FFP-simulations} extended these calculations with particular emphasis on FFP formation in both single and binary star systems. They find that a binary star is over 100 times more efficient in ejecting planets than planet-planet scatterings occurring in discs orbiting single stars. \cite{Coleman-24-FFPs} presented a detailed population study that takes into account FFP production by the CA scenario in both single and binary stars, finding that the binary channel dominates over the single star one strongly.

\cite{Coleman-24-FFPs} calculations yield $\sim 0.04$ JFFPs (mass grater than $0.3 \mj$), which is just a factor of $\sim 3$ below what is needed to account for $\mu$FFPs observations \citep{Mroz-23-Microlensing-planets-review,Yee-25-FFPs-MFunction}. However, there is a number of factors (such as considering Solar mass stars rather than M-dwarfs) that are not fully accounted for by \cite{Coleman-24-FFPs} which we believe reduce CA JFFP population significantly.

We now estimate an upper limit on the number of FFPs ejected by close M-dwarf binaries by assuming that gas giant planet formation in these systems is as efficient as in single star systems \footnote{This is a very optimistic assumption, given that binaries are expected to suppress planet formation due to chaotic dynamical environment \citep{PaardekooperEtal12,LinesEtal14,Marzari_19_CB_planets,Coleman-23-planet-in-CB-systems}.}. \cite{Clanton_16_Planet_mass_function} show that there is $N_{\rm p} \approx 0.06$ gas giant planets in this mass range per M-dwarf star. This is consistent with an earlier study by \cite{Bonfils_13_planet_MF} and Fig. 7 in \cite{Fulton_21_giants_occurence}.   The binary fraction of M-dwarf stars is $f_{\rm bin} \sim 0.25$ \citep{Offner_23_binaries_review,Clark_24_binaries}. Given the log-normal distribution for binary separations \citep{RaghavanEtal10,Offner_23_binaries_review} and assuming that close binaries responsible for FFP ejections have separation $0.05 < a_{\rm b} < 3$~au \citep{Coleman-24-FFPs}, we find that the fraction of such binaries if $f_{\rm close} \approx 0.2$. Further, CA predicts that massive solid core formation is unlikely in low metallicity environments \citep{IdaLin04b,IdaLin08,MordasiniEtal12}. This is consistent with observations that show that mean metallicity of giant planet-hosts is $ +0.2$ \citep{FischerValenti05}. At the same time, close binaries are known to anti-correlate strongly with metallicity of the stars, indicating that they probably formed by disc fragmentation \citep[e.g.,][]{Moe_Kratter_19_binary_vs_Z}. The mean metallicity of close binaries is $\approx -0.2$ \citep{Moe_Kratter_19_binary_vs_Z}. We therefore expect at least a factor of $f_{\rm Z} \sim 1/3$ suppression for efficiency of giant planet formation in close binaries. Finally, the fraction of planets ejected in simulations by \cite{Coleman-24-FFP-simulations} is $\sim 0.6$. Collating all of these factors together, we can expect that in the context of CA scenario,  close binaries will produce
\begin{equation}
    N_{\rm JFFP} < N_{\rm p}\times f_{\rm bin} f_{\rm close} f_{\rm Z} f_{\rm ej} = 0.009 \times N_{\rm p} \approx 6\times 10^{-4}\;
    \label{N_JFFP_CA}
\end{equation}
JFFP per star. On the other hand, in the mass range $(0.3-13) \mj$, there is $\approx 0.1$ FFPs per star in the microlensing surveys \citep{Gould-22-microlensingFFPs,sumi2023} that are identified in the planet-like formation channel \citep{Yee-25-FFPs-MFunction}. This is much larger  than the upper limit in eq. \ref{N_JFFP_CA}, and we note that we  assumed without definite proof that the observed gas giants in M-dwarfs are formed by Core Accretion. As shown by \cite{Liu_19_CA_low_Mstar,Miguel_20_CA_Mdwarfs,Mulders_21_CA_Mdwarfs} and others, CA faces significant difficulties in forming enough gas giants in M-dwarfs, and disc fragmentation has been suggested to fill the gap \citep{Bryant_25_giant_around_Mdwarf}. There thus appears to be a significant shortcoming of the JFFP fraction predicted by the  CA theory versus microlensing observations\footnote{Gavin Coleman, private communication, finds that preliminary calculations that include a stellar mass function down to 0.1$\msun$ and a realistic protoplanetary disc [Fe/H] distribution  do indeed reduce the number of JFFPs produced very strongly, in an approximate agreement with our estimate.}. On the other hand, observational uncertainties remain large. For example, the 1$\sigma$ uncertainties on the JFFP fraction in \cite{sumi2023} is a factor of $\sim 10$.

A comparison between the observed bound populations of gas giants and the microlensing JFFPs is another useful perspective. \cite{Zang_25_bound_microlensing} presents analysis of 63 KMTNET microlensing planet-mass events, concluding that there are $\sim 0.12$ gas giant planets (defined as an object with mass $0.3\mj \leq M_{\rm p}\leq 13\mj$) per star. The planets in the sample orbit their host stars at separations $a\sim (2-20)$~au. Assuming that most of these stars are single M-dwarf stars (recall that for M-dwarfs, the binary fraction is only $\sim 0.25$), the number of close binaries in the close binary regime ($a_{\rm bin}\lesssim 3$~au) per star in the \cite{Zang_25_bound_microlensing} population is expected to be $\sim f_{\rm bin} f_{\rm close} = 0.05$. If close binaries hatched planets as efficiently as single stars, and if all of these migrated in and were ejected, we would expect $\sim 0.12 \times 0.05 = 5\times 10^{-3}$ JFFP per star. Henceforth, for the close binary scenario \citep{Coleman-24-FFP-simulations} to eject the observed JFFPs,
we must assume that, at the epoch of planet formation, there existed a more numerous population of gas giants by a factor of $\sim 20$. This population would be present in protoplanetary discs of both single and binary stars. In that case, however, there should be many more gas giant planets on small orbits than observed, because in single stars these planets would not be ejected, and hence should still be there. As an example, fig. 30 in \cite{Clanton_16_Planet_mass_function} shows that there is only $\sim 0.05$ gas giant in the inner 1 au of a typical M-dwarf star.


Let us discuss the disc fragmentation scenario now. There are a few factors to consider:
\begin{enumerate}
    \item[a)] As we argued in \S \ref{sec:general_setup}, planets may form by disc fragmentation in the first $0.1$~Myr or less, before the secondary does. Thus, there is no physical reason for the secondary star to suppress planet formation in binaries in our framework. Note that forming a planet in a binary in our scenario does {\em not} imply it survives there. We argue that the planets form just as efficiently in binaries as in single star systems, but are very frequently ejected.
    \item[b)] Further to point (a), formation of a $q> 0.1$ secondary star by disc fragmentation is likely to require the most massive circum-primary discs simply to provide enough gas mass for the secondary to grow. More massive discs are also more likely to fragment to form planets \citep[as Toomre parameter $Q\propto M_*/M_{\rm disc}$, e.g., ][]{gammie2001}. Observations support these ideas as hosts of gas giant planets and brown dwarfs have wide binary fraction $\sim$ twice larger than non hosts  \citep{Fontanive_19_giants_in_wide_binaries}.  Similarly, simulations by \cite{Cadman_22_triggered_fragm_binary} show that secondary's presence on a wide orbit ($\gtrsim 200$~au) may in fact trigger disc fragmentation, helping, not hindering planet formation in binaries.
    \item[c)] Our scenario does not require close binaries (which are rare). We  start with a wide separation binary ($a_{\rm b} =140$~au), and end with $a_{\rm b}\sim$ tens of au, see Fig. \ref{fig:m2_a_mass}. In other words, our scenario may work for a ``garden variety" binary, although see \S \ref{sec:theory_implications_discussion}.
    \item[(d)] Disc fragmentation is expected to be more efficient at low metallicity stellar hosts since radiative cooling is more efficient \citep[e.g.,][]{gammie2001,Moe_Kratter_19_binary_vs_Z,Xu_25_RHD_disc_fragmentation}. Thus our scenario does not incur ``the metallicity penalty" factor, $f_Z$.
\end{enumerate}

Taking this all into account, let us estimate how many gas giant planets per star, $N_{\rm p, DF}$, is required to be born in the disc fragmentation scenario. For reasons discussed above, there is no $f_{\rm close}$ or $f_{\rm Z}$ factors. The ejection fractions we find are $\sim 0.5$ for $q\gtrsim 0.05$. If this can be generalized to a typical binary system, the required number of gas giant planets per star in the disc fragmentation scenario is
\begin{equation}
    N_{\rm p,DF} \sim \frac{0.12}{f_{\rm bin}f_{\rm ej}} = \frac{0.12}{0.25\times 0.5} \sim 1;
    \label{N_p_DF}
\end{equation}
The observed fraction of gas giants integrated over all separations is $\sim 0.2$ \citep[][]{Clanton_16_Planet_mass_function,Fulton_21_giants_occurence,Zang_25_bound_microlensing}. Further, at distances of tens of au,  direct imaging surveys \citep[e.g.,][]{BowlerEtal15} indicate a decrease in the giant planet abundance. The California Legacy Survey finds $\sim 0.09$ giant planets in the separation range of $8-32$~au. One could thus argue that there are an order of magnitude too few gas giant planets ``available" for FFP ejection in the first place.

If this was so, then no planet formation scenario in binary systems, whether CA or disc fragmentation, could ever account for the observed JFFPs. However,  in our model, the planets are on wide orbits of tens of au at $t\lesssim 0.1$ Myr, whereas direct imaging surveys probe systems older than $\sim 10$ Myr. In our disc fragmentation planet formation framework \citep[e.g., see a review in][]{Nayakshin_Review}, the number of gas giant planets {\em formed} at $t \ll 1$~Myr is always larger than the number of planets surviving beyond the disc dispersion age. The primary process that destroys post-collapse disc fragmentation planets is the catastrophic mass loss when the planet migrates closer in to the star, when it becomes sensitive to tidal disruptions \citep{BoleyEtal10,MachidaEtal11,cha2011}. These disruptions are in fact observable if/when the planet is disrupted in the inner $R \sim (0.05-0.1)$~au of the star. Focusing on episodic accretion onto young stars, \cite{VB05,VB06,VB10} have shown that in their 2D simulations most of gas clumps migrate from their birth places at tens to hundreds of au right to the inner boundary of their computational domain ($R\sim 10$~au). \cite{NayakshinLodato12} extended this modeling all the way to the stellar surface in their 1D disc simulations, finding that the planets are disrupted inside $\sim 0.1$~au when they migrate there. More recently, \cite{Nayakshin-23-FUOR,Nayakshin23-FUORi-2} have shown that the observations of FU Ori outbursts strongly favor this planet-disruption scenario. The statistics of FU Ori type outbursts shows that most protostars may go through $\sim 10$ of such accretion outbursts \citep{HK96,Contreras-19-FUOR-statistics,Contreras-Pena-24}, although models show that one planet may power multiple outbursts \citep{Nayakshin23-FUORi-2}. While the constraints from FU Ori outbursts remain patchy \citep[these outbursts are rare and their statistics remains an active area of research;][]{Fischer-PPVII}, it is likely that having $N_{\rm p,DF} \gtrsim  $ a few per disc is required to account for FU Ori outbursts statistics. The vast majority of these planets would need to be destroyed in the inner regions of protoplanetary discs rather than ejected\footnote{Compared with the CA scenario for giant planet formation, where gas giants are born at host star ages of $\sim 7$ Myr and with radii $\sim 2 \rj$ \citep{MordasiniEtal12a}, disc fragmentation planets are born at $t\sim 0.1-0.5$ Myr \citep{VB10,VB15}. They can be pushed all the way to the star much quicker since the discs are much more massive at that epoch, and they are also much more fluffy, having radii exceeding $10 \rj$ at young ages \citep[cf. ][]{Nayakshin-23-FUOR,Nayakshin23-FUORi-2}. For these reasons disc fragmentation planets are much more vulnerable to perishing in FU Ori outbursts than their CA-formed cousins.}.

On the other hand, disc fragmentation planets may open wide gaps in the disc, say at a few au, and do not migrate all the way to the star to be disrupted in FU Ori outbursts. If this is the case, then the only way to produce a sufficient number of JFFPs would be to have gas giant planet formation to be a factor of $\gtrsim 5$ more efficient in binaries than around single stars. This contradicts the basic principles of CA planet formation pathways \citep{Thebault_15_Stype_binaries}, but may be reasonable in the disc fragmentation picture where more massive discs may fragment on more objects.

\subsubsection{Velocity and cluster spatial distributions of FFPs}\label{sec:velocities}

\begin{figure}
    \centering
    \includegraphics[width=0.5\textwidth]{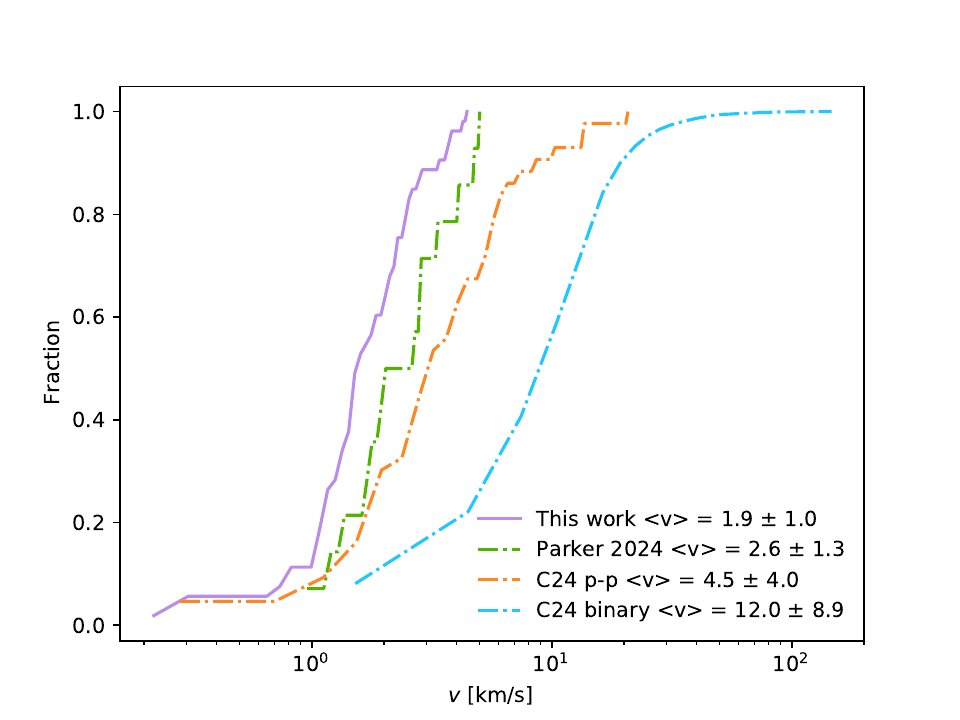}
    \caption{The cumulative distribution function for the ejected planet velocities (purple solid line), compared to ejected planets in simulations of a star forming region from \protect\cite{parker2024} (green dot dashed line) and planets ejected through planet-planet scattering (orange dot dashed line) and interactions with a central binary (blue dot dashed line) from \protect\cite{Coleman-24-FFP-simulations}. }
    \label{fig:parker}
\end{figure}

Fig. \ref{fig:parker} shows a cumulative distribution function for the ejected planet velocities in our simulations, compared to results obtained previously by \cite{parker2024} and \cite{Coleman-24-FFP-simulations}. \cite{parker2024} used N-body simulations to study production of FFPs via stellar flybys in young star cluster NGC 1333 \citep{Luhman-16-NGC1333}. As mentioned earlier, \cite{Coleman-24-FFP-simulations} studied planetary ejections in close binary systems in the CA scenario.

The final velocities of ejected planets we obtain in this paper is low, i.e., $v_{\rm f} = (1.9\pm 1.0)$ km s$^{-1}$ (see also Fig. \ref{fig:vf_slice}). Our ejection velocities are in agreement with previous work by \cite{Stamatellos_07_BDs,Basu_Vorobyov_12_BDs}. These values are lower but comparable to that obtained by \cite{parker2024} (the green dot-dash curve in Fig. \ref{fig:parker}). This is to be expected as their planets (see their Table 1) are placed on initial orbits with semi-major axis of 30 au, i.e., only marginally smaller orbits than that of our planets, so only a slightly larger kick is needed (cf. Fig. \ref{fig:vkick}) to eject their planets compared to ours. In spite of the seemingly small difference in the final velocity of the ejected planets between the simulations presented here and those of \cite{parker2024}, observations may be able to distinguish between these scenarios via differences in the spatial distribution of JFFPs and host stars. \cite{parker2024} finds (cf. their Fig. 4) that since their JFFPs are on average faster than field stars, they are distributed more widely spatially in the simulated star cluster NGC1333 compared with the stars. Our JFFPs are slower, and so may be distributed more like their host stars. Setting up our simulations in a star-cluster environment is necessary to investigate this question in greater detail.

\cite{Coleman-24-FFP-simulations} find that their planet-planet interactions (the yellow curve in Fig. \ref{fig:parker}) produce ejections with velocities on average a few times higher than ours. However, planet-planet ejections are exceedingly rare in his simulations, as \cite{Coleman-24-FFPs} find that the binary-planet ejection channel (the blue curve in Fig. \ref{fig:parker}) is by far the dominant one. This is especially true for the massive JFFPs that we study in this paper \citep[see Fig. 2 in][where the binary-planet ejections are at least an order of magnitude more likely than planet-planet ejections]{Coleman-24-FFPs}. The binary planet ejection channel of \cite{Coleman-24-FFP-simulations,Coleman-24-FFPs} produces JFFPs that are on average an order of magnitude faster than our ejectees. This is natural given that in the context of CA scenario, planets are kicked from very small orbits $a\sim 0.5$ to a few au, so a much larger velocity kick is needed to eject them. 

We conclude from this that planets ejected by binaries from GI discs have the lowest velocities when compared to ejections by stellar flybys and planets ejected in the CA scenario. Future observations which give us the velocity dispersions of FFPs may therefore shed light not only on the ejection mechanism of FFPs, but also on their planet formation pathway. Even in the absence of direct velocity measurements of JFFPs, significant constraints on their velocity distribution could be placed in young stellar clusters. Consider an NGC1333-like stellar cluster with a physical size of $R\sim 1$~pc. The velocity dispersion in the cluster is of order $\sim 2$ km/sec \citep{parker2024}, thus a planet ejected from a star with excess velocity of $v_{\rm f}\sim 10$~km/sec is unbound with respect to the cluster. Its residence time in the cluster is
\begin{equation}
    t_{\rm res} = \frac{R}{v_{\rm f}} \approx 0.1 \;\text{Myr}\;,
\end{equation}
which is a small fraction of the cluster age. Hence  CA  \citep{Coleman-24-FFP-simulations,Coleman-24-FFPs} produces {\em runaway} JFFPs for this cluster, and hence finding a significant population of JFFPs still present in the cluster would be a very strong argument in favor of producing them via GI. For example, assuming that CA scenario ejects 1 planet per compact binary system over $t_{\rm ejection} \sim 1$ Myr time scale \citep[e.g., Fig. 3 in][]{Coleman-24-FFP-simulations}, we estimate that the fraction of CA JFFPs still inside the cluster would be
\begin{equation}
    f_{\rm present} \sim \frac{t_{\rm res}}{t_{\rm ejection}} \sim 0.1\;.
\end{equation}
Coupled with the general inefficiency of CA in producing JFFPs (\S \ref{sec:numbers_discussion}), this makes it exceedingly unlikely that any JFFPs observed in young clusters could come from this scenario.

\subsubsection{The ages of JFFPs}\label{sec:ages_discussion}


CA binaries eject gas giants after $\sim 1$ Myr of binary formation \citep[Fig. 3 in][]{Coleman-24-FFP-simulations}. However, these calculations were performed for Solar metallicity, Solar type stars. As shown by \cite{MordasiniEtal12}, formation of gas giant planets occurs much later (if at all) in low metallicity systems, such as the close binary population \citep{Moe_Kratter_19_binary_vs_Z}. It would be very interesting to consider the JFFP ejection ages in the CA scenario for a realistic stellar mass and metallicity distribution. It may well be that JFFP ejections would occur significantly later in this case. In contrast, our disc fragmentation scenario results in JFFP ejection at very young system age, $t \sim 0.1$~Myr. JFFPs are observed in young stellar clusters, with their  ages  from $t\sim 1$ to $\sim 10$ Myrs \citep[e.g.,][]{Lucas2001-FFP-Orion, Luhman-16-NGC1333, miret-roig2022,Seo_Scholz_25}. It appears that some of these JFFPs are too young for the CA scenario.

\subsection{Theoretical Implications and future work}\label{sec:theory_implications_discussion}

Despite being the first detailed planet formation theory to appear \citep{Kuiper51b,McCreaWilliams65}, the full life-cycle of the disc fragmentation scenario has only began to emerge recently \citep{BoleyEtal10,nayakshin2010,Nayakshin_Review}, when the role of planet migration in massive self-gravitating discs was finally appreciated \citep{VB06,VB10,MachidaEtal11,baruteau2011,ForganRice13b}. With all the recent progress \citep[as reviewed by, e.g.,][]{KratterL16}, a number of unresolved issues remain, such as the role of dust and solids in forming gas giant planets by disc fragmentation \citep[e.g.,][]{RiceEtal04,GibbonsEtal14,Baehr_22_GI_dust,Baehr-23-GI-dust,Longarini_23_solids_collapse_theory,Longarini_23_solids_collapse_sims}, realistic turbulence and magnetic fields \citep{Deng20-MHD-GI,Deng_21_GI_Neptunes,Kubli-GI-MHD-23}. Excitingly, JFFPs offer a new set of observables to constrain the disc fragmentation scenario. 

The most significant limitation of our simulations is not following up the actual formation of the planets and the secondary; we instead explored dynamical consequences of having Jupiter-mass planets in the disc in which a seed of the secondary formed. In the future, it is desirable to relax these assumptions.

We stop our simulations at $\sim 10^5$ years due to finite computing time resources. It is pertinent to ask whether following up the system for longer would not yield more JFFP ejections. This is quite likely, and we aim to address it in future simulations. 

We do not take into account stellar flybys that are likely to take place on longer time scales \citep{Lin_98_FFPs,Vorobyov_17_FFP_flyby,parker2024,yu2024,Fu_25_FFP_from_flybys}; this may well eject more of the survivors into the field. However, to eject a significant number of planets,  fly-bys need to occur very early on, e.g., well within the first 0.1 Myr of planet formation by GI, since simulations of planet migration in GI discs show that they migrate inward on the time scale as short as $3-10$ thousand years \citep{machida2010,BoleyEtal10,baruteau2011}. Most of these planets would be lost to the inner disc by the time the fly-by occurs, so that, to unbind the planet, the fly-by must have a very small (a few au) impact parameter, which is statistically quite unlikely.

\section{Conclusions}\label{sec:conclusions}

We presented 3D hydrodynamical simulations of Jupiter mass planets, and a more massive seed of a secondary star on a wider orbit, embedded in a self-gravitating disc around $M_*=0.5\msun$  star. The seed grows by accretion and is expected to become a low mass star. The secondary migrates inward in the disc rapidly, sweeping through the orbits of the planets. Our main conclusions are:

\begin{itemize}
    \item [(i)] Rapid migration of the growing secondary star leads to frequent orbital crossings that give significant velocity kicks to the planets. The kicks are stochastic and may put some planets on smaller orbits (e.g., see Fig. \ref{fig:close_encounter_59}). However, a series of such ``random walk" velocity kicks will usually terminate in an ejection of the planet (e.g., planets 2 and 3 in Fig. \ref{fig:planet7}). The physics of these ejections is related to the well-known gravitational assist effect (\S \ref{sec:toy}).
    
    \item[(ii)] We find the ejection fraction to increase with secondary mass, $M_{\rm s}$, strongly, growing from $\sim 20$\% at  $M_{\rm s}\approx 5 \mj$ to $\sim 70$\% at $M_{\rm s}\approx 50\mj$ (Fig. \ref{fig:ejection_frac_mass}). The ejected planets must be in the dense post-collapse configuration, or else they are disrupted rather than ejected (\S \ref{sec:disruptions}).

    \item[(iii)] The distance of the planet-secondary closest approach vary from a fraction of an au to $\sim 10$~au (Figs. \ref{fig:beta20_m210mj}--\ref{fig:beta30_m250mj}). This agrees with analytical estimates for a velocity kick sufficient for planet ejection (\S \ref{sec:toy}). This shows that most pre-collapse (gas clump-like planets) would be disrupted rather be ejected by the planet-secondary interactions, whereas most post-collapse gas giant planets (those with radii less than $\sim 20 \rj$) would survive the interactions intact. Observations of JFFPs could therefore set new independent constraints on the physics of massive fragmenting discs: only models in which most surviving giant planets born in the disc migrate towards the star and are disrupted there, can satisfy observational constraints (\S \ref{sec:numbers_discussion}).

    \item[(iv)]  Analytical arguments suggest that JFFPs ejected in our simulations may be able to retain discs from a small fraction to at most $\sim 2$~au in radius (\S \ref{sec:disc_retention}). 
    
    
    \item[(v)] Disc fragmentation and Core Accretion FFP formation channels can be distinguished via planet excess (ejection) velocities and their residence time in the cluster (\S \ref{sec:velocities}). Disc fragmentation FFPs have low final velocities, i.e., $v_{\rm f}\sim 2 \pm$ 1 km s$^{-1}$, which are lower than the post-ejection velocities of the other ejection mechanisms (cf. Fig. \ref{fig:parker}). Due to this, a good fraction of disc fragmentation JFFPs are likely to remain in their parent clusters for Myrs after ejection. In contrast, CA circumbinary planet ejections have an order of magnitude higher final velocities. Such JFFPs escape parent clusters in a small fraction of a Myr and cannot carry away any significant protoplanetary discs. 
    
    
    
    \item[(vi)] The mechanism presented in this paper produces JFFPs very early on. This is significant, as many JFFPs are found in young star forming regions, with ages from $t \sim$ 1 to a few Myrs \citep[e.g.,][]{Lucas2001-FFP-Orion, Luhman-16-NGC1333, miret-roig2022}. CA theories are expected to hatch giant planets at ages exceeding these time scales \citep{MordasiniEtal12,johansen2017-CAreview}.
\end{itemize}

\section*{Acknowledgements}

We acknowledge useful comments on this paper from Kevin Luhman, Aleks Scholz, Eduard Vorobyov, Phil Lucas, and Fabo Feng. AC is supported by STFC PhD studentship. NMR acknowledges support from the Beatriu de Pinós postdoctoral program under the Ministry of Research and Universities of the Government of Catalonia (Grant Reference No. 2023 BP 00215). This research used the ALICE High Performance Computing facility at the University of Leicester, and the DiRAC Data Intensive service at Leicester, operated by the University of Leicester IT Services, which forms part of the STFC DiRAC HPC Facility (\href{www.dirac.ac.uk}{www.dirac.ac.uk}). 

\section{Data availability}

The data obtained in our simulations can be made available on reasonable request to the corresponding author. The software used to run and visualise the simulations, \textsc{phantom} and \textsc{splash} are publicly availible from \href{https://github.com/danieljprice/phantom}{https://github.com/danieljprice/phantom} and \href{https://github.com/danieljprice/splash}{https://github.com/danieljprice/splash}
respectively.

\appendix
\section{Planet numbers per disc.}\label{sec:Appendix_planet_number}

\begin{figure}
    \centering
    \includegraphics[width=0.4\textwidth]{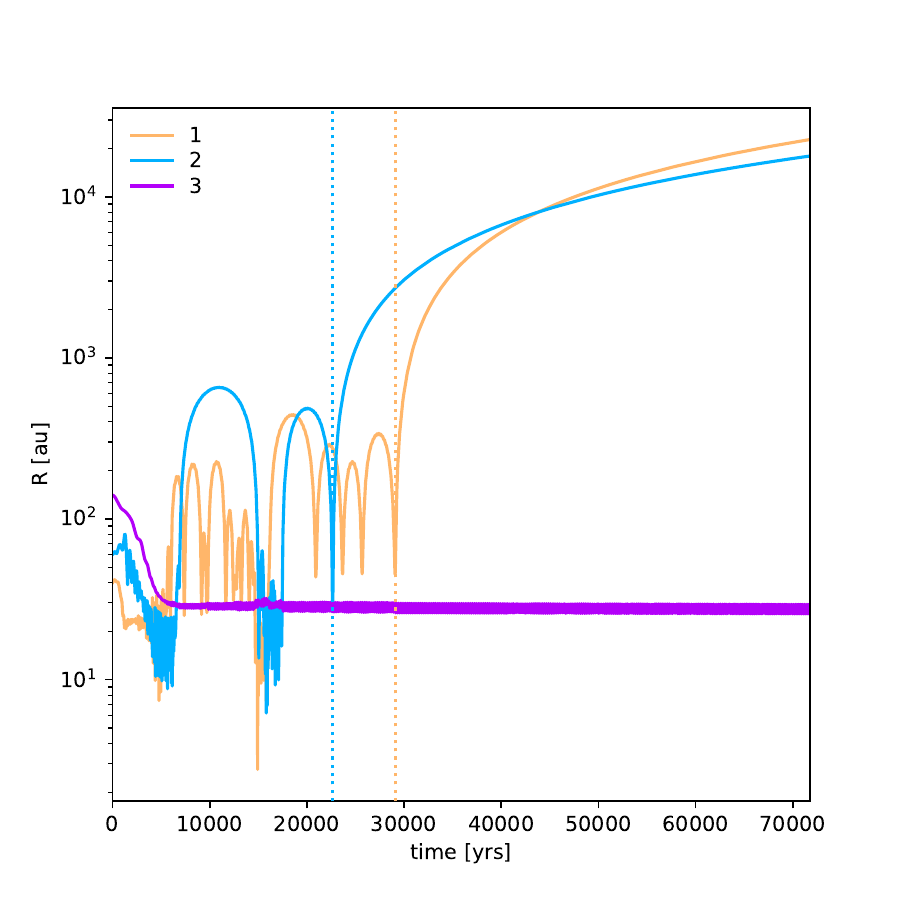}
    \caption{A two planet per disc experiment. The planets of $M_{\rm p}=1\mj$ mass are started on orbits with initial radii of 40 and 60 au, respectively.}
    \label{fig:2planets_r_40_60}
\end{figure}

One may suspect that using 8 planets per disc unduly influences our results, by, e.g., planet-planet scatterings ejecting some of the planets. This is not so for many reasons. First, the middle panels in Figs. \ref{fig:beta20_m210mj}--\ref{fig:beta30_m250mj} show planet-secondary distance. Whenever an ejection occurs, this distance shrinks to a minimum, showing that it is the close planet-secondary interaction that ejected the planet, not planet-planet interactions. Second, simulations that had no secondary (only planet mass objects) ejected no planets (Fig. \ref{fig:ejection_frac_mass}). Third, we ran three additional simulations with 4 rather than 8 planets, and four more with 2 planets per disc, all with a $M_{\rm 2} = $ 25 $\mj$ secondary object (one such experiment is shown in Fig. \ref{fig:2planets_r_40_60}). The average planet ejection fraction in these simulations is 0.61, very much consistent with Fig. \ref{fig:ejection_frac_mass}. Fourth, Nayakshin et al 2025, to be subm., present simulations with Phanton, and two additional codes, 2D fixed grid FARGO-ADSG \citep{Baruteau_Masset_08_radiative_disc} and N-body code REBOUND \citep{Rein_12_Rebound}, to investigate our FFP formation model further. They explore a wider range of initial conditions, and find a similarly high ejection fraction with the three different codes. They also consider particles in the test-mass regime, finding similarly high ejection fractions, which also shows that planet-planet interactions are at most a small perturbation to the main effect -- planet-secondary star interactions.

\section{Robustness of the ejection fraction}\label{sec:Appendix_ejection_fraction}

All of our SPH simulations started with the same planets and secondary object configuration (cf. Table \ref{tab:inital}). One may worry that the high ejection fraction we obtain is not representative of the broader parameter space.  Here we address this question with \texttt{REBOUND} \citep{Rein_12_Rebound}, an N-body code, which does not include hydrodynamics of the disc. We model the secondary star's migration with a simple exponential drag law:
\begin{equation}
    \frac{\mathrm{d}{\bf v}}{\mathrm{d}t} = - \frac{{\bf v}}{\tau_0}\;,
    \label{dvdt}
\end{equation}
where $\tau_0=10$~kyr is a free parameter. In realistic protoplanetary discs, massive objects open deep gaps and their migration slows down to a, relatively speaking, crawl \citep[Figs. \ref{fig:single_tracks}, \ref{fig:m2_a_mass}, and also][]{Malik2015}. Therefore we also introduce a secondary migration stopping distance, $a_{\rm stop}$, setting $\tau_0=\infty$ for time $t > \tau_0 \ln(a_0/a_{\rm stop})$, where $a_0$ is the secondary's initial semi-major axis. Despite its simplicity, this model includes the most important ingredient of our model, the planet-secondary star interactions. Furthermore, the \texttt{REBOUND}'s `ias15' N-body integrator \citep{Rein_15_Rebound_ias} betters \texttt{Phantom} in resolving very-close N-body interactions, allowing one to track bodily collisions between particles. The execution time of a \texttt{REBOUND} simulation is many orders of mangitude shorter than our SPH runs, permitting rapid exploration of parameter space. 

\begin{figure*}
    \centering
    \includegraphics[width=0.26\textwidth]{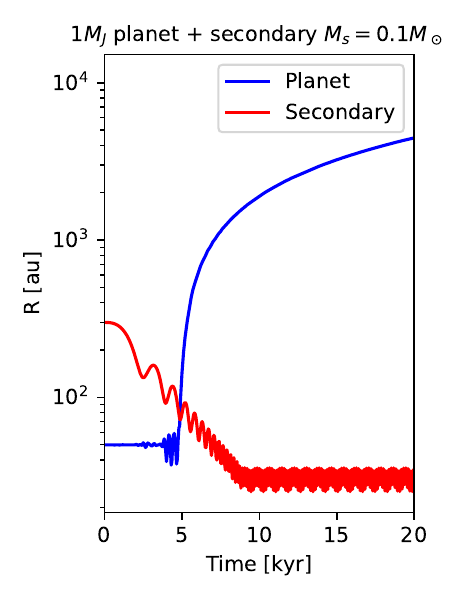}
    \includegraphics[width=0.35\textwidth]{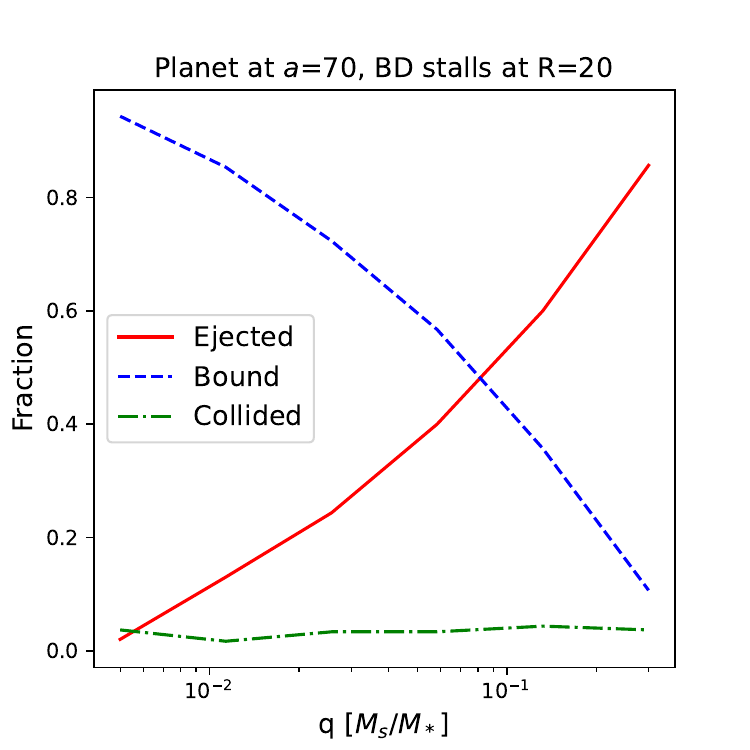}    \includegraphics[width=0.35\textwidth]{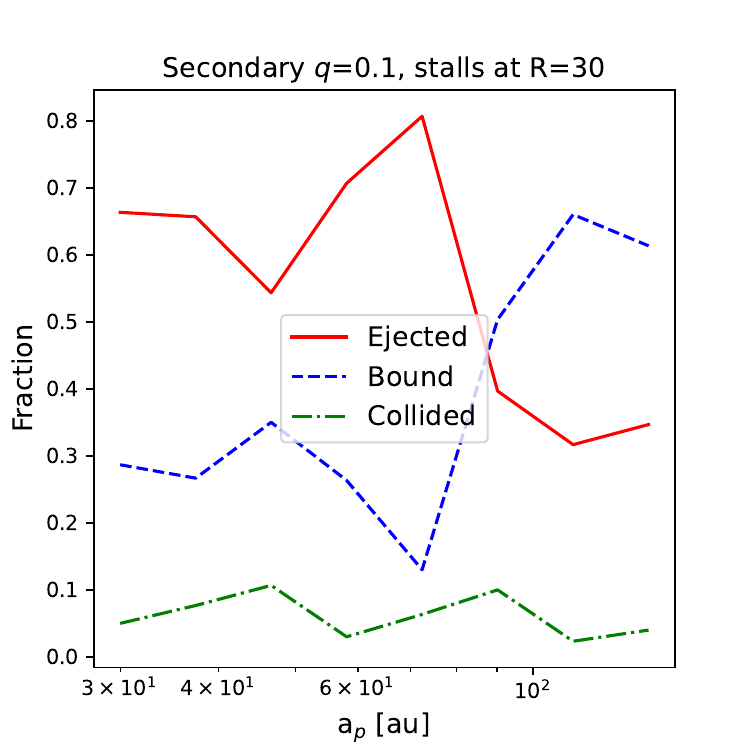}
    \caption{N-body simulations performed with \texttt{REBOUND}. Left: An example single planet run ending in an ejection. Middle: Fractions of planets ejected, remaining bound, and collided with the secondary star in similar experiments as a function of $q$. The planet is initialised at $a=70$~au. Right: Same but for a fixed $q$ and varying planet initial location. Note the similarities of these results with Figs. \ref{fig:ejection_frac_mass} and \ref{fig:sinkIDhist}.}
    \label{fig:Rebound}
\end{figure*}

Here we present a limited statistical exploration of the parameter space, and we share the \texttt{REBOUND} Jupyter Notebooks for these (see \href{https://github.com/sn6985/Rebound_FFPs/tree/main}{https://github.com/sn6985/Rebound\_FFPs/tree/main}), so that the reader can repeat and expand these experiments if desired. We run 300 experiments for each planet and secondary initial orbital configuration. Unlike our \texttt{Phantom} runs, here the initial azimuthal position of the planet and the secondary are uniform randoms. Fig. \ref{fig:Rebound} presents one example run in the left panel. The middle and the right panels show the fractions of planets ejected, remaining bound, or collided with the secondary star (we set secondary's radius to be $1 R_\odot$) for two contrasting assumptions. In the middle panel, the initial orbital positions of the planet and the secondary are constant, $a_{\rm p}=70$ and $a_0 = 300$ au, respectively, and we vary the secondary's mass from $0.005 \msun$ to $0.3\msun$. In the right panel of Fig. \ref{fig:Rebound}, on the other hand, we set $q=0.1$, $a_0=150$ and $a_{\rm stall}=30$~au, but we vary $a_{\rm p}$ from 30 to 140 au.

We observe that the collision fraction is always small, i.e., a few \% in the middle panel, reaching at most $\sim 10$\% for $a_{\rm p}\gtrsim a_{\rm stop}$ in the right panel. The ejection fraction increases strongly with $q$, reaching $\sim 0.8$ for $q=0.3$. This corroborates Fig. \ref{fig:ejection_frac_mass}, although there are quantitative differences, such as the absence in the drop at $M_{\rm s}=15 \mj$. This is due to our migration model (eq. \ref{dvdt}) being too simple compared with Fig. \ref{fig:single_tracks}, which shows that objects with intermediate mass are those migrating inward the fastest, hence having less time to affect the planets initially at $R \gtrsim 60$~au.

The right panel of Fig. \ref{fig:Rebound} bears a remarkable similarity to Fig. \ref{fig:sinkIDhist}, showing indeed that these are the planets on orbits around $a_{\rm stall}$ that suffer the violent ejections (or bodily collisions) most frequently. The planets that are initially close to the secondary, on the other hand, tend to remain bound (although their semi-major axis may increase) because the secondary passes their vicinity quickly. The final orbits of these planets are usually stable according to the \cite{Holman_Wiegert_99} criteria. The peak in the ejection fraction at $a_{\rm p}\sim 80$~au is due to those planets being affected by the secondary both during its migration and at its final stalling location.

These experiments corroborate the rather large ejection fraction that we find in our paper, $f_{\rm ej}\gtrsim 0.5$. In Nayakshin et al (2025), different initial conditions, and three different codes (\texttt{Phantom}, \texttt{REBOUND} and \texttt{FARGO-ADSG} \citep{Baruteau_Masset_08_radiative_disc,Marzari_12_circumbinary_discs}), are used, and the ejection fraction is found to be similarly high. We feel certain that our scenario can be very efficient in ejecting planets.

\bibliographystyle{mnras}
\bibliography{ref.bib,nayakshin.bib} 







\bsp	
\label{lastpage}
\end{document}